\documentclass[12pt]{article}
\usepackage{amsmath}
\usepackage{graphicx,psfrag,epsf}
\usepackage{enumerate}
\usepackage[hyphens]{url}
\usepackage[natbibapa]{apacite}
\usepackage{setspace}

\usepackage{xcolor}
\usepackage[margin=1in]{geometry}

\newcommand{\blind}{1}

\date{\vspace{-5ex}}

\begin{document}

\def\spacingset#1{\renewcommand{\baselinestretch}%
{#1}\small\normalsize} \spacingset{1}


\if1\blind
{
  \title{\bf Multiple system estimation using covariates having missing values and measurement error: estimating the size of the M\=aori population in New Zealand}
  \author{Peter G.M. van der Heijden\thanks{Utrecht University, Padualaan 14, 3584 CH Utrecht, The Netherlands and University of Southampton, Highfield, Southampton, SO17 1BJ, UK. E-mail: {p.g.m.vanderheijden@uu.nl}},
Maarten Cruyff\thanks{Utrecht University, Padualaan 14, 3584 CH Utrecht, The Netherlands. E-mail: {m.cruyff@uu.nl}},
Paul A. Smith\thanks{University of Southampton, Highfield, Southampton, SO17 1BJ, UK. E-mail: {p.a.smith@soton.ac.uk}},\\
Christine Bycroft\thanks{Statistics New Zealand. E-mail: christine.bycroft@stats.govt.nz},
Patrick Graham\thanks{Statistics New Zealand. E-mail: patrick.graham@stats.govt.nz},
Nathaniel Matheson-Dunning\thanks{Statistics New Zealand. E-mail: nathaniel.matheson-dunning@stats.govt.nz}\\}
  \maketitle
} \fi

\if0\blind
{
  \begin{center}
    {\LARGE\bf Multiple system estimation using covariates having missing values and measurement error: estimating the size of the M\=aori population in New Zealand}\\
\hfill \break

\end{center}
  
} \fi
\newpage
\begin{abstract}
We investigate use of two or more linked registers, or lists, for both population size estimation and to investigate the relationship between variables appearing on all or only some registers. This relationship is usually not fully known because some individuals appear in only some registers, and some are not in any register. These two problems have been solved simultaneously using the EM algorithm. We extend this approach to estimate the size of the indigenous Māori population in New Zealand, leading to several innovations: (1) the approach is extended to four registers (including the population census), where the reporting of Māori status differs between registers; (2) some individuals in one or more registers have missing ethnicity, and we adapt the approach to handle this additional missingness; (3) some registers cover subsets of the population by design. We discuss under which assumptions such structural undercoverage can be ignored and provide a general result; (4) we treat the Māori indicator in each register as a variable measured with error, and embed a latent class model in the multiple system estimation to estimate the population size of a latent variable, interpreted as the true Māori status. Finally, we discuss estimating the Māori population size from administrative data only. Supplementary materials for our article are available online.
\end{abstract}

\noindent%
{\it Keywords: capture-recapture, population size estimation, latent class model, register coverage, x}
\vfill

\newpage
\spacingset{1.5} 
\section{Introduction}
\label{sec:intro}
The use of dual system estimation (DSE, also known as capture--recapture or the Lincoln--Peterson estimator) to estimate the size of a population which cannot be completely observed has become widespread in official statistics, particularly as a key part of making estimates from a population census (e.g. Wolter 1986, Brown et al. 1999, 2019)\nocite{BrownJJDiamondIDChambersRL1999}\nocite{BrownJSextonCAbbottOSmithPA2019}\nocite{WolterKM1986}, though also in situations involving the use of linked administrative data sources (Bakker et al., 2015,\nocite{BakkerBFMvanderHeijdenPGMScholtusS2015} and references therein; Zhang \& Dunne, 2018\nocite{ZhangDunne2018}). The need to make efficient use of data already available to government in the construction of official statistics outputs has led to better access to administrative data, but there are important challenges in making use of it (see, for example, Hand, 2018)\nocite{Hand2018}. Linkage of the records from these sources is being widely used to understand the under-- and over--coverage within them and estimate coverage corrections. We will use “registers” as a generic term for all sources containing lists of identifiable units.

When two registers are linked, in general there will be some records which remain unlinked, as there is no corresponding record in the other source. This leads to missing data for any variables which appear in only one register (item missingness). The linked data can be used to estimate the size of the population that is not present in either register, and for these unobserved records \textit{all} the variables are missing (unit missingness). There is an extensive line of research that treats this problem from a missing data perspective, starting with Zwane and van der Heijden (2007)\nocite{ZwaneEvanderHeijdenPGM2007}, and summarized and extended in van der Heijden et al. (2012, 2018)\nocite{VanderHeijdenPGMWhittakerCruyffBakkerVanderVliet2012}\nocite{VanderHeijdenPGMSmithPCruyffMBakkerB2018}. De Waal, van Delden and Scholtus (in press, see also 2019)\nocite{deWaalvanDeldenScholtus2019}\nocite{deWaalvanDeldenScholtus2020} provide an overview of what they call multi--source statistics, where they categorize this line of research under the subject ``Overlapping variables and overlapping units'' and ``Undercoverage''. Another overview of the field is provided by Zhang and Chambers (2019)\nocite{ZhangChambers2019}. The missing data methodology involves the EM algorithm, using a suitable loglinear model specified by the researcher. In the EM algorithm an Expectation step and a Maximization step are alternated until convergence. In the E--step, the expected values of the missing data are derived given the observed data and the current best fitted values found in the preceding M--step. In the M--step, maximum likelihood estimates are found for the chosen model using the completed data from the E--step; see Zwane \& van der Heijden (2007) for the full details.
Standard errors for the estimates can be calculated using the parametric bootstrap (Buckland and Garthwaite, 1991)\nocite{BucklandSGarthwaiteP1991}.
Van der Heijden et al. (2018) concluded that further practical experience with these methods is needed to demonstrate their usefulness in a variety of situations and encouraged their wider application.

Here we consider these methods for estimating the size of the M\=aori ethnic population in New Zealand, and section \ref{maori} describes the context of this problem, the registers available and the procedures which have been used to link them. Some of these registers cover only parts of the population, a common situation in official statistics applications of population size estimation. For the estimation of ethnicity in New Zealand this part coverage is related to the age of individuals. Zwane, van der Pal-de Bruin and van der Heijden (2004)\nocite{ZwaneEvanderPaldeBruinKvanderHeijdenPGM2004} approached the problem of registers that do not fully cover the population as a missing data problem. They provided solutions assuming the missingness (i.e. the undercoverage) is ignorable, which is often not a priori  unrealistic as the missingness is due to the design of the registers (in contrast to other missing data problems where ignorability \textit{is} often unrealistic). In the current paper we reframe the part--coverage of a register as a collapsibility problem for a covariate. Using results of van der Heijden et al. (2012), we show under what assumptions we may assume collapsibility over such a covariate, and hence, when the part--coverage of registers may be ignored. This is discussed in section \ref{collapse}.

Then we build up the estimation problem in section \ref{build}, starting with two registers, then progressing to three and four registers.
In section \ref{nocensus} we consider estimates obtained only from the administrative registers, to examine how good estimates would be in the absence of a population census. Section \ref{measurement error} introduces the idea that a different concept of M\=aori identity is measured in each register because of differences in context or timing, an extension of the idea that the same variable is measured differently in different registers (van der Heijden et al. 2018). The model is extended to include latent classes to represent an underlying concept of M\=aori/non-M\=aori. Under this model a single estimate is obtained for the size of the M\=aori and non-M\=aori populations (instead of estimates for each register separately). Section \ref{discussion} concludes by discussing the effectiveness of these approaches and the corresponding estimates of the size of the M\=aori ethnic population, and how the differences in the estimates derived from the different registers may be interpreted. It also discusses the sensitivity of the approaches to assumptions of perfect linkage and no overcoverage.

\section{M\=aori ethnicity and relevant registers}
\label{maori}
Ethnicity is the principal measure of cultural identity in New Zealand, and is used across the official statistics system. Identifying the indigenous M\=aori population is of particular importance due to the partnership and obligations between M\=aori and the crown established under the Treaty of Waitangi of 1840. M\=aori are also a key group of policy interest; for example Waldon (2019)\nocite{WaldonJ2019} discusses the way measurement of ethnicity supports the measurement of health outcomes for indigenous peoples in New Zealand.

Ethnicity is therefore regularly included in statistical and administrative data collections. However, differences in questions or context and changes in perceptions can all affect how ethnicity is recorded in different registers (Simpson et al. 2016)\nocite{SimpsonLJivrajSWarrenJ2016}. Particularly for indigenous peoples such as the M\=aori, there is a need to ensure that definitions are created which take into account the ways in which members of these communities perceive themselves (Madden et al. 2019)\nocite{MaddenRColemanCMashfordAConnolyM2019}.

Official population estimates and projections for major ethnic groups in New Zealand are based principally on the responses people provide in the five yearly census, adjusted for nonresponse using a post--enumeration survey. As part of its census transformation programme, Statistics New Zealand is exploring the feasibility of a census based on administrative data (Statistics New Zealand 2012, 2014) \nocite{StatisticsNewZealand2012}\nocite{StatisticsNewZealand2014}. The ability to produce ethnicity data from administrative registers is a key consideration. Ethnicity is collected independently in a number of administrative registers as well as through the census. People do not always report the same ethnicity in each register and sometimes do not report their ethnicity at all, so there is an additional missingness problem to deal with.

A key question is how to combine ethnicity from multiple registers, when information is sometimes conflicting. Reid, Bycroft, and Gleisner (2016) \nocite{ReidBycroftGleisner2016}
compared ethnicity data from the 2013 Census with the ethnicity information collected by administrative registers, for a New Zealand resident population derived from administrative registers. They found that nearly everyone in this administrative data--based New Zealand resident population had ethnicity recorded in at least one administrative register, but that consistency with census responses varied considerably by register and by ethnic group. The method used to combine these registers has a major impact on the result. Under the assumption that census responses provide the best measure for official statistics purposes, a method that ranks registers based on their consistency with the census was applied. Using administrative data alone was found to produce a time series that reflects expected patterns of increasing ethnic diversity, with the age structure and regional distribution of ethnicity consistently in line with official measures (Statistics New Zealand, 2018)\nocite{StatisticsNewZealand2018}.
We note that, according to Statistics New Zealand's standard classification of ethnicity, used in many administrative systems, people can and do provide multiple ethnicities.
Here we focus on M\=aori ethnicity so that we have two mutually exclusive categories:  M\=aori (with or without other ethnicities) and non--M\=aori (everyone else).

Four registers are available, the population census and three administrative registers, each with an ethnicity variable. In this application we use ethnicity information from the linked administrative data to explore alternative ways to produce official ethnic population estimates in New Zealand, and also investigate estimates produced from only the administrative data as a possible replacement for the census. In support of this we analyse a variety of census and administrative registers using the approach of Zwane and van der Heijden (2007), with a specific focus on the estimation of the size of the M\=aori population at the time of the 2013 population census. The analysis requires the extension of the methods to deal with multiple registers and with a variety of different types of missing data.

The population used here is the experimental administrative--based New Zealand resident population known as the ‘IDI-ERP’ (Statistics New Zealand, 2017a)\nocite{StatisticsNewZealand2017a}. The IDI-ERP is derived using signs of activity in government sources. Those who have died, or who have moved to live overseas before the reference date are excluded to minimise over-coverage, though some non-residents may remain in the dataset. We are interested in the population size, and therefore will implicitly assess the coverage of different sources within IDI-ERP relative to our estimate of the size of the New Zealand population.

The data are probabilistically linked in Stats NZ’s Integrated Data Infrastructure (IDI). The IDI provides safe access to anonymised linked microdata for research and statistics in the public interest. Data sources in the IDI (including the census) are linked to a central population spine. Perfect linkage is an essential assumption for DSE. An incorrect link could mean that the wrong ethnicity is associated with a person. In this application, if records in the registers have not been linked to the IDI spine, they do not enter the analysis, and become part of the unobserved population for the register.

The three administrative registers are:

\begin{itemize}
\item Department of Internal Affairs (DIA) birth registrations data -- which includes the ethnicity of the child as reported at registration
\item Ministry of Education (MOE) tertiary education enrolment data -- which includes ethnicity for students
\item Ministry of Health (MOH) National Health Index system, a unified national person list -- which includes ethnicity
\end{itemize}
For a more detailed explanation of these registers, see Reid et al. (2016).

Each of the administrative registers relates to different parts of the population. Birth registrations are for babies born in NZ since 1998, or those up to age 14 in 2013; tertiary education enrolments are available from the late 1990s, and are mainly for those aged between 18 and 40 years in 2013; both census and health data include all ages, and each register has an ethnicity value for around 90 \% of the IDI-ERP population. Overall, almost 99 \% of the IDI-ERP population have ethnicity information from at least one of these registers, and many people have information from more than one register. Table \ref{marg} provides the observed counts for ethnicity, where $-$ stands for item missingness (individuals that do not have their ethnicity registered in this register) and x stands for individuals that are not part of a register. For example, in the Census 3,225,804 are registered as non--M\=aori, 560,427 as M\=aori, for 20,619 individuals in the census no ethnicity is reported, and 595,140 individuals are missed by the census but appear in at least one of the other registers.

\linespread{1.00}
\begin{table}[h!]
\bf \caption{\rm \it Summary of Census linked to DIA, MOH and MOE, observed numbers}
\label{marg}\rm
\begin{center}
\begin{tabular}{lrrrr}
\hline
              & Census    & DIA       & MOH         & MOE \\
\hline
non--M\=aori  & 3,225,804 & 574,077   & 3,527,874 & 1,763,463\\
M\=aori       & 560,427	  & 236,673   & 617,205	  & 405,063\\
$-$           & 20,619	  & 6,045     & 188,781	  & 20,424\\
x             & 595,140   &	3,585,195 &	68,130	  & 2,213,040\\
\hline
TOTAL         & 4,401,990 & 4,401,990 & 4,401,990 & 4,401,990\\
\end{tabular}
\end{center}
\end{table}
\linespread{2.0}

The aim is to produce aggregate estimates of M\=aori and non--M\=aori ethnicity by combining these four independent registers: the 2013 Census and the three administrative registers. In the next section, we first ask the question whether it is a problem that the four registers cover different parts of the population.

\section{Using registers that cover different parts of the population}
\label{collapse}

Usually expositions about multiple system estimation assume that each register aims to cover the same population. That is not the case here, where the Census and MOH aim to cover the complete population but DIA and MOE aim to cover only subpopulations. Building on earlier work of Zwane et al. (2004)\nocite{ZwaneEvanderPaldeBruinKvanderHeijdenPGM2004} and van der Heijden et al. (2012)\nocite{VanderHeijdenPGMWhittakerCruyffBakkerVanderVliet2012} we will argue here that ignoring the fact that DIA and MOE only cover a subpopulation is not problematic for estimating the size of the population. We devote some space to this problem as it is one that is regularly encountered in population size estimation. An alternative modelling approach is presented by Sutherland and Schwarz (2007)\nocite{SutherlandJMSchwarzCJRivestLP2007}, and related work is found in di Cecco et al. (2018), van der Heijden and Smith (2020)\nocite{heijden2020estimating}, and di Cecco, di Zio and Liseo (2020)\nocite{diCeccodiZioLiseo2020}.

\subsection{Partial coverage and collapsibility, two registers}
\begin{figure}
\begin{center}
  \includegraphics[width=\textwidth]{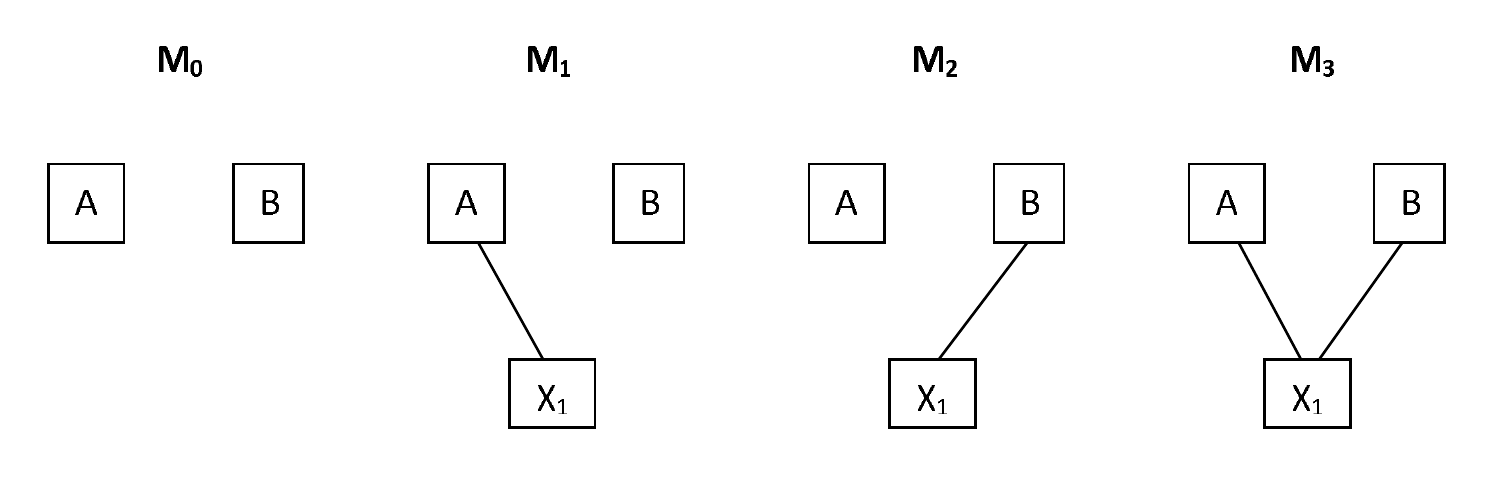}
  \caption{Interaction graphs for loglinear models with two registers and one covariate, taken from van der Heijden at al. (2012)}\label{Fig1}
\end{center}
\end{figure}

We now reframe the problem. We show how these two results are related to collapsibility in loglinear population size estimation models described by van der Heijden et al. (2012). See Figure \ref{Fig1}, taken from their paper. In all models there are two registers, $A$ and $B$. In model $M_0$ there is no covariate. In model $M_1$ there is a covariate $X_1$ that is related to register $A$ but not to register $B$, meaning that inclusion probabilities are heterogeneous across the levels of $X_1$ for $A$ but homogeneous for $B$. In model $M_2$ it is the other way around. In model $M_3$ inclusion probabilities for both $A$ and $B$ are heterogeneous across the levels of $X_1$. Van der Heijden et al. show that the total population size estimate is identical in models $M_0$, $M_1$ and $M_2$ and different from the estimate in $M_3$. In other words, under models $M_1$ and $M_2$ the three-way array of variables $A, B$ and $X_1$ is collapsible over covariate $X_1$. Under model $M_3$ there is no collapsibility. They use the concept of a \textit{short path} to distinguish situations where there is collapsibility from situations where there is not. A short path is a sequence of connected nodes in the graph which does not contain a sub-path; note that it need not be short in the sense of having few nodes. For full details see Whittaker (1990)\nocite{WhittakerJ1990} and van der Heijden et al. (2012). In $M_3$ the covariate lies on a short path between $A$ and $B$ and therefore one cannot collapse over $X_1$. In $M_1$ and $M_2$ the covariate $X_1$ does not lie on a short path and in these cases one can collapse over the covariate.

Now we reframe the results from Zwane et al. (2004) in terms of the results of van der Heijden et al.. In Example 1 there are two regions: register A covers the full population, but register B covers only the north. Here, region can be considered as the covariate $X_1$. If we assume that the inclusion probabilities for $A$ are homogeneous over region, there is no edge from $X_1$ to $A$. Yet the inclusion probabilities for $B$ are heterogenous, as these are positive for the north and zero for the south region, and there is an edge from $X_1$ to $B$. By the assumption of homogeneous inclusion probabilities for $A$, region is not on a short path between $A$ and $B$. Hence we can collapse over region -- in other words, ignore the problem that register $B$ only covers the north region. This is the graph in model $M_2$ in Fig. \ref{Fig1}. Example 2 has three regions, with A covering the north and central regions, and B covering the central and south regions. This situation is described by model $M_3$: $A$ has heterogeneous inclusion probabilities over the levels of region as the inclusion probabilities are positive for the north and the middle but zero for the south, and and so does $B$ as the inclusion probabilities are zero in the north and positive in the middle and south. Hence there are two edges, one from $X_1$ to $A$ and another from $X_1$ to $B$, and therefore $X_1$ \textit{is} on a short path from $A$ to $B$ and it follows that we cannot collapse over (i.e. ignore) the region covariate.

\subsection{Extension to more than two registers}

\begin{figure}
\begin{center}
  \includegraphics[width=\textwidth]{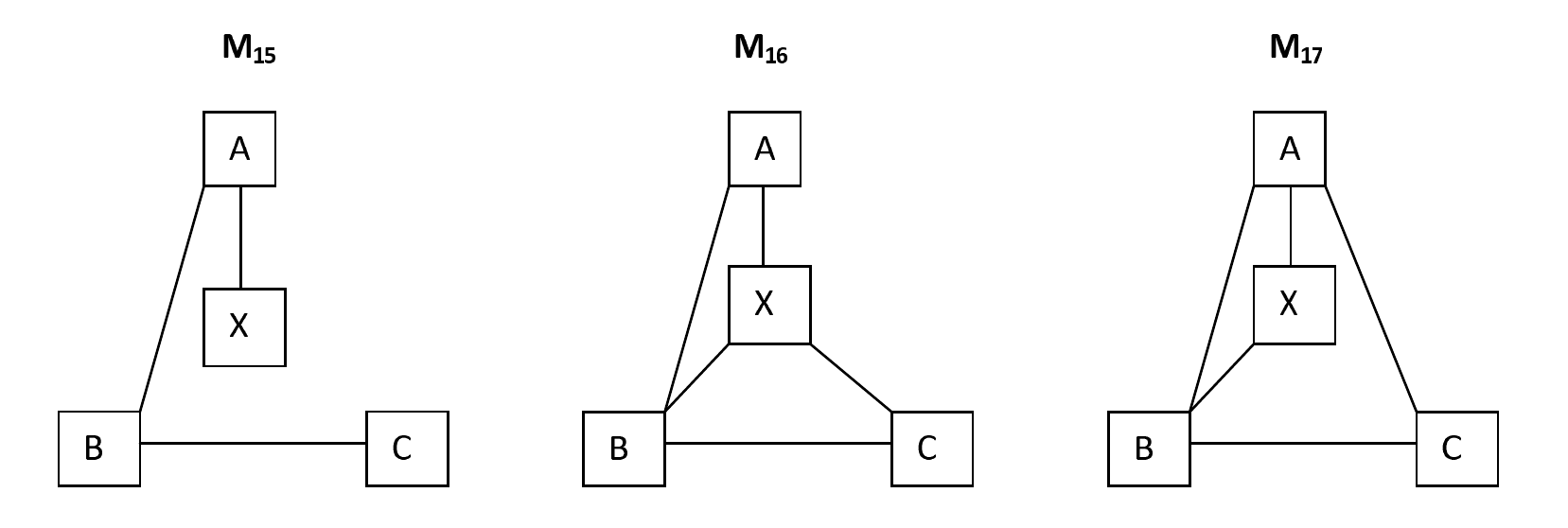}
  \caption{Interaction graphs for loglinear models with three registers and one covariate, taken from van der Heijden at al. (2012)}
  \label{Fig4new}
\end{center}
\end{figure}

Using the short path concept van der Heijden et al. (2012) provide the following results for three registers, see Figure \ref{Fig4new}, reproduced from their paper. In model $M_{15}$ we can collapse over the covariate $X$ as it is not on a short path. In model $M_{16}$ we cannot collapse over $X$ as it is on the short path between registers $A$ and $C$. In model $M_{17}$ we can collapse over $X$ as $X$ is not on a short path from $A$ to $B$. The maximal model, $[ABX][ACX][BCX]$ (using the notation that the interaction within [ ]'s and all lower order interactions involving the same terms are included), not shown in Figure \ref{Fig4new}, is not collapsible over $X$.

We apply these results in the next sections, when we produce estimates based on linking three and four registers. At this stage of the paper we cannot immediately discuss the models used for the estimation of ethnic background. In these models there are not only variables for ``being in register 1,...,4'' but also covariates ``ethnicity in register 1,...,4'' which makes the models more complicated than the models discussed in this section. In Section \ref{build} we start with a model for registers Census and MOH, that both aim to cover the full population. Then we add register DIA, the birth registration started in 1998. Conceptually, we consider there to be a covariate Age, for which we do not have data: inclusion probabilities are high for individuals born from 1998 onwards and zero before that time. Here we have a model similar to $M_{15}$, as the inclusion probabilities for DIA are related to Age but for the Census and MOH we assume they are not. When we add register MOE we assume for this register that there is an unmeasured variable that is also related to age, as enrolments are available starting in the late 1990s, and that there are other factors that lead one to go into tertiary education. For Age we end up with a model having properties similar to those of model $M_{17}$, as Age is related to two registers but if there is also a direct link between the two registers the model is collapsible over Age. The other factors leading one to go to tertiary education are further covariates that are only related to MOE and not to the other registers and the model is therefore collapsible over these covariates too. We come back to this in the next sections.

\subsection{General result on the ignorability of partial coverage}
We can extend the approach of this section to give a general result. For any number of registers, and any combination of full and partial coverages of the population, the partial coverage will be ignorable (that is, we can obtain the correct results from modelling without taking any special account of the population coverages) if the model is collapsible over the covariate (or covariates) which defines the partial coverage(s). Equivalently, the variable(s) defining the coverage(s) do not appear on any short path in the graphical representation of the model.

\section{Building up the population estimates}
\label{build}

\subsection{Two registers}
\label{2reg}
We start by using the two registers with the widest coverage, the Census and the MOH. Being in the Census is denoted by $A$ ($A = 1$ for `yes', $A = 0$ for `no'), and similarly for MOH, denoted by $C$. The ethnicity variable in the Census is denoted by $a$ ($a = 0$ for non--M\=aori, $a = 1$ for M\=aori, $a = -$ for individuals who are in $A$ but did not fill in their ethnicity, and $a$ = x for individuals that are not in $A$). The ethnicity variable in the MOH is denoted by $c$ and coded similarly to $a$. In comparison to the methods employed by van der Heijden et al. (2018)\nocite{VanderHeijdenPGMSmithPCruyffMBakkerB2018}, the presence of the $-$ level in variables $a$ and $c$ is new, and we first extend the approach to deal with this new level with two registers.

\begin{figure}
\begin{center}
  \includegraphics[width=\textwidth]{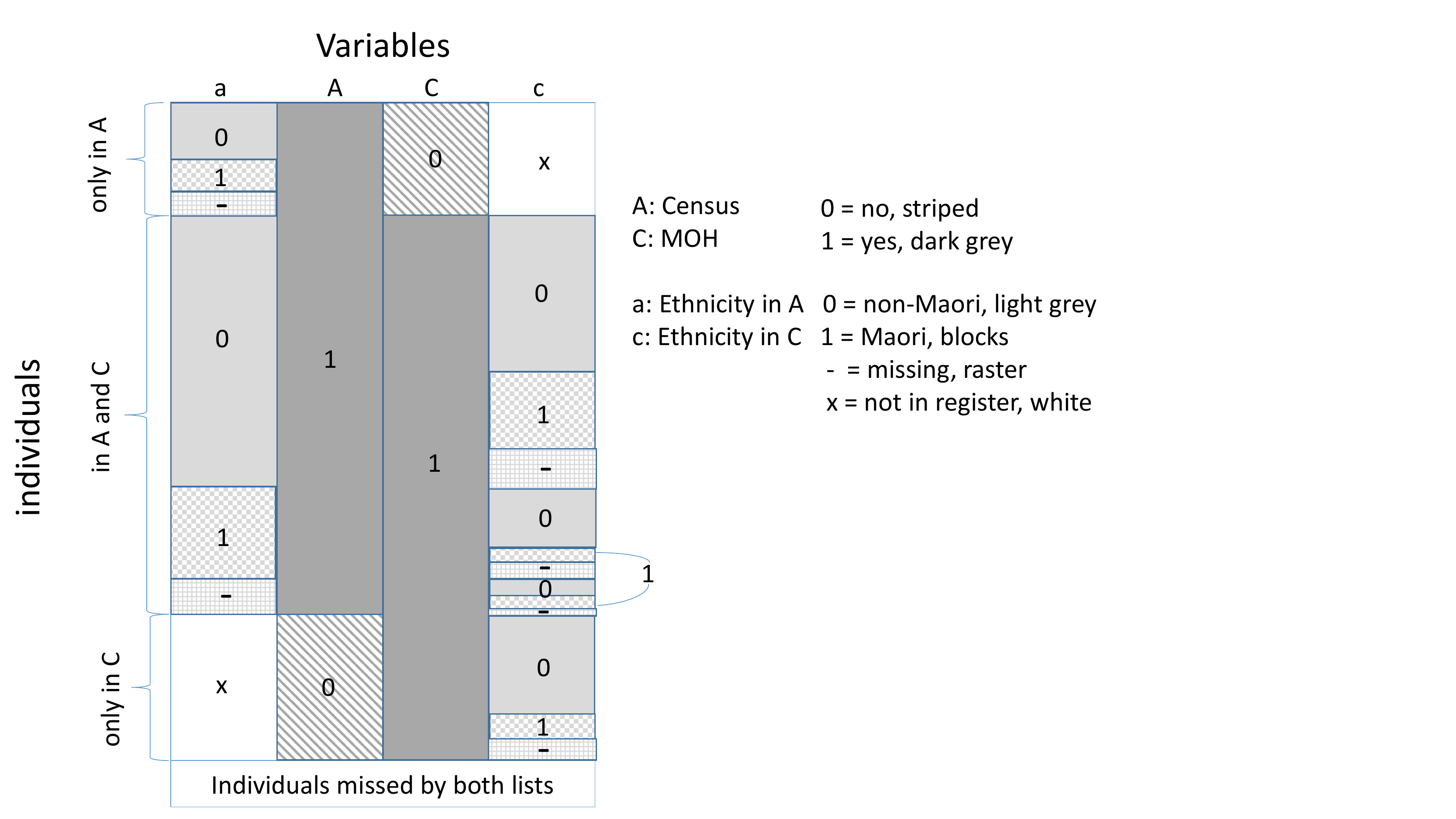}
  \caption{Graphical representation of two linked registers}
  \label{Figure1}
\end{center}
\end{figure}

Figure~\ref{Figure1} illustrates the form of the data when they are coded in a matrix of individuals in the rows by variables in the columns. In the middle two columns we depict $A$ and $C$, that indicate whether individuals are only in $A$ but not in $C$ ($(A, C) = (1, 0)$), in both $A$ and $C$ ($(A, C) = (1, 1)$) or not in $A$ but only in $C$ ($(A, C) = (0, 1)$). At the bottom we find a horizontal band of `Individuals missed by both lists', and this refers to $(A, C) = (0, 0)$. This last number has to be estimated to arrive at an estimate of the size of the total population of non--M\=aori and M\=aori. The first column stands for ethnicity variable $a$. When individuals are in $A$ ($(A, C) = (1, 0)$ or $(A, C) = (1, 1)$), there are three types of individuals, namely 0, non--M\=aori (light grey); 1, M\=aori (checkerboard pattern); and $-$, those who have a missing value for ethnicity (gridded pattern). When individuals are not in $A$ but only in $C$, the ethnicity variable $a$ is automatically not measured and denoted by x (white area). The last column stands for ethnicity variable $c$, and it has similar levels to $a$. Notice that there are three kinds of missing data: there is item missingness $-$ for those individuals that are on a list but did not provide their ethnicity; there is item missingness x for those individuals that are not on one list, and hence have no value on the corresponding ethnicity variable (if only $A = 0$, $a =\textrm{x}$, and if only $C = 0$, $c =\textrm{x}$). Last, there is unit missingness for those individuals that are in neither $A$ nor $C$.

The problem can also be presented in contingency table format, see Table \ref{missing}, panel 1. In the rows we have the combination of variables $A$ and $a$, with levels $(A, a) = (1, 0), (1, 1), (1, -), \newline (0, \textrm{x})$, and similarly for the combination of variables $C$ and $c$ in the columns. Thus there is a $4 \times 4$ table with a complicated structure. The $3 \times 3$ subtable top--left shows the cross--classification for individuals in both $A$ and $C$. The first two diagonal values are very large, showing that many individuals have the same ethnicity in both the census and MOH. The counts 108,189 and 31,995 refer to the number of individuals for which the information on ethnicity in $a$ contradicts that in $c$. The counts for $a = -$ show the number of individuals that are in $A$ but whose ethnicity is missing. The aim of our analysis will be to distribute (for example) these 16,512 individuals over the levels $a = 0$ and $a = 1$. It is clear that, even though most of these individuals will be non--M\=aori ($a = 0$) as they are non--M\=aori in $c$, not all of them are non--M\=aori as for non--M\=aori $c = 0$ there are still 108,189 individuals that are classified as M\=aori in $a$ ($a = 1$). Similarly for the counts for $c = -$. The 900 individuals for which the variables $a$ and $c$ are both missing ($a, c) = (-, -)$ will have to be distributed over the four cells $(a, c) = (0, 0), (0, 1), (1, 0), (1, 1)$. When individuals are not in the Census, $A = 0$, then $a$ is automatically missing. In the cross--classification with $C$ and $c$, the numbers of individuals with $A = 0$ and $a = \textrm{x}$  have to be distributed over $A = 0, a = 0$ and $A = 0, a = 1$.

\linespread{1.0}
\begin{table}[h!]
\bf \caption{\rm\textit{Census ($A$) linked to MOH ($C$). Ethnicity in $A$ is denoted by $a$ and ethnicity in $C$ is denoted by $c$, where $a$ and $c$ have levels \rm 0 \it (non--M\=aori), \rm 1 \it (M\=aori), $-$ (missing) and \rm x \it (not in register). The data have been randomly rounded to base 3 to protect confidentiality. Source: Stats NZ}.}
\label{missing}\rm
\begin{center}
\begin{tabular}{ll rr rr r}
\multicolumn{7}{l}{\it Panel 1: Observed counts}\\
\\ \hline
&&\multicolumn{3}{c}{$C=1$}&\multicolumn{1}{c}{$C=0$}&Totals\\
&&$c=0$&$c=1$&$c= -$&          \multicolumn{1}{c}{$c=$ x}&\\
\hline
\\
$A=1$ &$a=0$&   3,004,335  &  31,995  & 150,840 &\multicolumn{1}{r}{ 38,634}&3,225,804   \\
      &$a=1$&     108,189  & 435,465  &  12,405 &\multicolumn{1}{r}{  4,368}&  560,427   \\
      &$a=-$&      16,512  &   2,769  &     900 &\multicolumn{1}{r}{    438}&   20,619 \\
\\
$A=0$ &$a=$x&     398,838  & 146,976  &  24,636 &\multicolumn{1}{r}{-}      &  570,450 \\
\\
Totals&     &   3,527,874  & 617,205	& 188,781 &\multicolumn{1}{r}{ 43,440}&4,377,300\\
\hline
\\
\\
\multicolumn{7}{l}{\it Panel 2: Fitted values under  $[Ac][ac][Ca]$}\\
\\ \hline
&&\multicolumn{2}{c}{$C=1$}&\multicolumn{2}{c}{$C=0$}&Totals\\
&&$c=0$&$c=1$&$c=0$&$c=1$ & \\
\hline\\
$A=1$  &$a=0$ & 3,170,294.8    &    33,787.9 &  38,616.0  &  411.6 &3,243,110.3\\
      &$a=1$ &   111,242.5    &   448,084.8 &     877.6  &3,534.9 &  563,739.8\\
\\
$A=0$  & $a=0$&  402,709.4     &    10,770.8 &   4,905.2  &  131.2 &  418,516.6\\
      & $a=1$&   14,130.7     &   142,839.1 &     111.5  &1,126.8 &  158,208.1\\
\\
Totals&       & 3,698,377.4    &   635,482.6 &	44,510.3  &5,204.5 &4,383,574.8\\
\hline
\end{tabular}
\end{center}
\end{table}


\linespread{2.0}

The original 15 counts in Table \ref{missing}, Panel 1, will have to be redistributed over 3 subtables of dimension $2 \times 2$. I.e., the subtable of size $3 \times 3$ has to be reduced to size $2 \times 2$, the three values for $A = 0, a = \textrm{x}$ have to lead to a subtable of size $2 \times 2$ and similarly for the three values for $C = 0, c = \textrm{x}$. In a second step the subtable for $A = 0, C = 0$ has to be estimated, and this refers to the individuals that are missed by both lists. Thus two types of missing data are estimated.

For the simpler situation that there are no missing data of type $-$ but only of type x, van der Heijden et al. (2018) describe how this result can be found using an EM algorithm with some loglinear model specified by the researcher. Van der Heijden et al. (2018) show that the maximal loglinear model that can be fitted to the data is $[Ac][ac][Ca]$, where the highest order fitted margins are placed between square brackets. If we denote the levels of $A, C, a, c$ by $R, T, r, t$, then this model can be denoted by
\begin{equation}
\label{LL-model}
\log \pi_{RTrt} = \lambda + \lambda^A_R + \lambda^C_T + \lambda^{a}_r + \lambda^{c}_t + \lambda^{Ac}_{Rt} + \lambda^{Ca}_{Tr} + \lambda^{ac}_{rt},
\end{equation}
with identifying restrictions that the parameters $\lambda$, $\lambda^A_1,$ $\lambda^C_1$, $\lambda^{a}_1,$ $\lambda^{c}_1,$ $\lambda^{Ac}_{11},$ $\lambda^{Ca}_{11}$ and $\lambda^{ac}_{11}$ are free, and the other parameters are restricted to be zero. The maximal model $[Ac][ac][Ca]$ is saturated in the sense that the fitted values are equal to the observed values. The other two--factor interactions cannot be estimated as there are no data to estimate them, for example, the interaction between $A$ and $C$ cannot be estimated as the joint marginal count for $A = 0$ and $C = 0$ is zero, and similarly for $A$ and $a$, and for $C$ and $c$.

The EM algorithm can also be applied in this more complicated situation with missing data of type $-$ and x with both types replaced by their expected values in the E-step. The result is given in Table \ref{missing}, Panel 2. We created our own code (see suppl. materials B and C) instead of the computer program CAT that we used in the past (Schafer, Harding and Tusell, 2012; van der Heijden et al., 2012, supplementary materials). Due to the fitted model, in each of the three estimated $2 \times 2$ subtables the  odds ratio between $a$ and $c$ is identical and equal to 377.9, the value from the observed top--left subtable.

The lower right $2 \times 2$ table in Panel 2 of Table \ref{missing} shows the estimated numbers of people missing from both census and MOH. These numbers are relatively low, due to the large overlap of the two registers. Under independence between $A$ and $C$ conditional on $a$ and $c$, we find, for $a=0$ and $c=0$, the estimate $4,905.2 = 38,616.0 \times 402,709.4 / 3,170,294.8$, and this low estimate is due to the large denominator (where $A=1$ and $C=1$). Although not many individuals are missed by both registers, approximately one fifth of the missed individuals are M\=aori.

The parameter estimates are given in Table \ref{parest2}. Their standard errors are very small, which is not surprising given the large observed frequencies. Notice that the estimate 4,905.2 in panel 2 of Table \ref{missing} can also be obtained as exp(8.4981). The estimated conditional odds ratio between the ethnicity variables $a$ and $c$ of 377.9 can be obtained as exp(5.9347). Also notice that the relation between $A$ and $c$ is negative, showing that being in the census, $A$, goes together with a smaller probability of being recorded as M\=aori in the MOH (estimated odds ratio is exp$(-0.9201) = 0.40$), whereas being in MOH, $C$, goes together with a larger probability of being recorded as M\=aori in the census (estimated odds ratio is exp(0.4344) = 1.54).

\linespread{1.0}
\begin{table}[h!]
\bf \caption{\rm \it Parameter estimates for two registers model}
\label{parest2}\rm
\begin{center}
\begin{tabular}{lrrrr}
\hline
            &Estimate   &Std. Error& z value & Pr\\
\hline
(Intercept) &  8.50 &     NA&        NA &  NA      \\
A    &        2.06 & 0.002&  1248.1  &  $ <0.001$\\
c    &      $-3.62$& 0.006& $-585.2$ &  $ <0.001$\\
C    &        4.41 & 0.005&   865.1  &  $ <0.001$\\
a    &      $-3.78$& 0.016& $-233.7$ &  $ <0.001$\\
A:c  &      $-0.92$& 0.003& $-268.9$ &  $ <0.001$\\
C:a  &        0.43 & 0.016&    27.1  &  $ <0.001$\\
c:a  &        5.94 & 0.007&   903.7  &  $ <0.001$\\
\hline
\end{tabular}
\end{center}
\end{table}
\linespread{2.0}

To estimate confidence intervals we use a procedure that can be considered, conceptually, to be a hybrid between the non-parametric and the parametric bootstrap. The nonparametric bootstrap draws random samples of size $n$ from the frequencies observed in the sample (including in our example the frequencies of – and x, that is from Table 2, Panel 1). This approach takes the uncertainty due to these missing values into account, but not the uncertainty due to the estimated population size. As a consequence, the nonparametric bootstrap tends to underestimate the variance of $\hat{N}$ (Buckland and Garthwaite 1991, Anan, Bohning and Maruotti, 2017). The parametric bootstrap draws random samples of size $\hat{N}$ from the fitted frequencies (in our example proportions derived from Table 2, Panel 2), and then sets the counts for cells which would not be observed (e.g. the cells for which $A=0$ and $C=0$) to zero.  This approach, however, does not generate observations of missing values for the ethnicity variables $a$ and $c$, and therefore also underestimates the variance of $\hat{N}$ because it excludes the imputation of these missing values.

Therefore we use a new, hybrid bootstrap procedure here that has the aim to combine these two properties: we use the observed counts in Panel 1 and supplement these with the fitted value of the total of the subtable for $A = 0, C = 0$ in Panel 2. These observed counts and the $A = 0, C = 0$ fitted value are used to derive multinomial probabilities (using $\hat{N}$ as the denominator). Bootstrap samples of size $\hat{N}$ (rounded) are then drawn with these probabilities, after which the counts in the subtable for $A = 0, C = 0$ are set to zero, and the new data analysed with model  \ref{LL-model}. In this application 2,000 bootstrap samples are used to derive a confidence interval using the percentile method.

The estimated total population size for New Zealand is 4,383,575; the 95 percent confidence interval using the percentile method, estimated with the parametric bootstrap, is 4,383,404 -- 4,383,736. The census and the MOH differ in which part of this total is M\=aori, as summarised in Table \ref{summary}, Panel A. The variable measuring ethnicity with the best validity can be used. If there is no clear preference, a practical solution could be to average the two estimates. Other considerations are discussed in section \ref{4reg} below. When the number of registers involved is larger than two, we propose to use latent class models to bring the separate estimates into agreement with each other, see Section \ref{measurement error}.

\linespread{1.0}

\begin{table}[h!]
\bf \caption{\rm\textit{Summary of population size estimates of non-M\=aori and M\=aori under the selected models according to the ethnicity classifications in different registers, estimated numbers and 95 percent confidence intervals}.}
\label{summary}\rm
\begin{center}
\begin{tabular}{l c c c c c c}
\hline
      & \multicolumn{3}{c}{M\=aori}&\multicolumn{3}{c}{Non-M\=aori}\\ \hline
      & estimate & 2.5 percent & 97.5 percent & estimate & 2.5 percent & 97.5 percent\\ \hline
      \\
\multicolumn{7}{l}{\it Panel A: Two registers (section 4.1)}\\ \hline
Census & $721,948$ & $720,499$ & $723,542$ & $3,661,627$ & $3,660,031$ & $3,663,081$\\
MOH    & $640,687$ & $639,226$ & $642,233$ & $3,742,888$ & $3,741,370$ & $3,744,338$ \\ \hline

\\
\multicolumn{7}{l}{\it Panel B: Three registers (section 4.2)}\\ \hline
Census & $729,123$ & $727,440$ & $730,822$ & $3,690,122$ & $3,686,382$ & $3,694,110$ \\
DIA    & $771,217$ & $768,867$ & $773,608$ & $3,648,027$ & $3,643,997$ & $3,652,166$ \\
MOH    & $642,724$ & $641,129$ & $644,307$ & $3,776,521$ & $3,772,866$ & $3,780,461$ \\ \hline

\\
\multicolumn{7}{l}{\it Panel C: Four registers, restricted model (section 4.3)}\\ \hline
Census & $733,294$ & $731,608$ & $734,947$ & $3,689,668$ & $3,687,698$ & $3,691,559$ \\
DIA    & $775,697$ & $772,236$ & $779,109$ & $3,647,265$ & $3,643,429$ & $3,651,285$ \\
MOH    & $645,112$ & $643,533$ & $646,707$ & $3,777,849$ & $3,775,988$ & $3,779,663$ \\
MOE    & $762,222$ & $760,103$ & $764,323$ & $3,660,740$ & $3,658,421$ & $3,662,854$ \\ \hline

\\
\multicolumn{7}{l}{\it Panel D: Three registers, administrative data sources only (section 4.4)}\\ \hline
DIA & $804,936$ & $800,904$ & $809,185$ & $3,600,293$ & $3,595,293$ & $3,605,915$ \\
MOH & $641,495$ & $639,936$ & $643,043$ & $3,763,734$ & $3,760,042$ & $3,768,416$ \\
MOE & $780,234$ & $777,752$ & $782,557$ & $3,624,995$ & $3,620,873$ & $3,630,055$ \\ \hline

\end{tabular}
\end{center}
\end{table}


\subsection{Three registers}
\label{3reg}
We now add the DIA register, denoted $B$, with ethnicity variable $b$, and consider data derived from three registers. The observed counts are found in the online supplementary material. Adding a third register has the advantage that we no longer have to assume conditional independence of $A$ and $C$ as in the model $[Ac][ac][Ca]$. Now we can fit models that have pairwise dependence between the three registers, which makes the fitted model more realistic. The cells of the contingency table with the observed frequencies are denoted with $A,B,C \in \{0, 1\}$ and $a,b,c \in \{0, 1, -, $x$\}$, and the cells of the contingency table with the fitted frequencies with $A,B,C,a,b,c \in \{0, 1\}$, since the missing levels $-$ and x have been imputed.

\begin{figure}
\begin{center}
  \includegraphics[width=\textwidth]{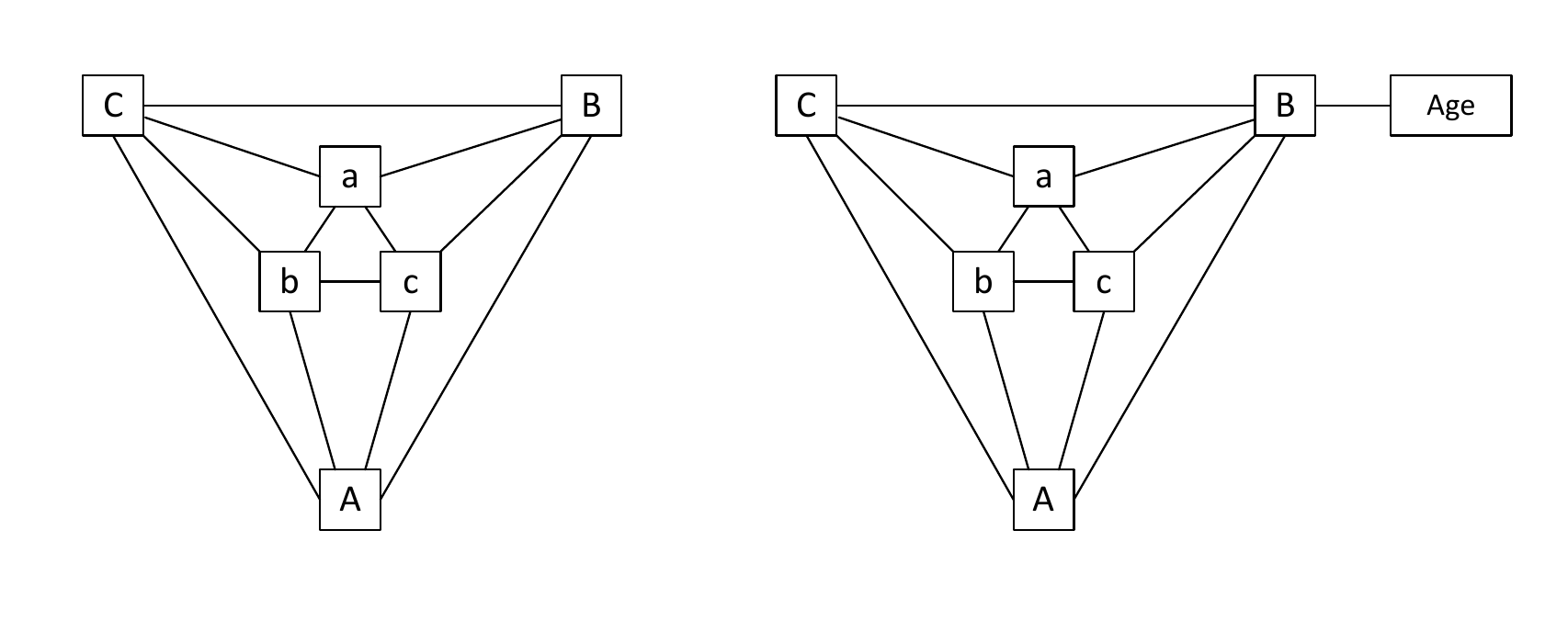}
  \caption{Interaction graphs for loglinear model $[ABc][ACb][BCa][Abc][Bac][Cab][abc]$ on the left, extended with a covariate Age on the right}
  \label{Graph3}
\end{center}
\end{figure}

As there are 6 variables, the potential number of fitted frequencies is $2^6 = 64$. However, there are 8 structurally zero cells for the combination of $A = 0, B = 0$ and $C = 0$. Also, as in the situation for two registers, it is not possible to fit models where $A$ interacts with $a$, $B$ with $b$ or $C$ with $c$. Thus the maximal model is $[ABc][ACb][BCa][Abc][Bac][Cab][abc]$. This model has 1 (intercept) + 6 (main effects) + 12 (15 two--factor interactions minus the three interactions that cannot be fitted) + 7 (three factor interactions) = 26 parameters. See the left graph in Figure \ref{Graph3} for a graphical model. As described in Section \ref{collapse}, it is clear that DIA ($B$) has heterogeneous inclusion probabilities over age, as for some age categories the inclusion probabilities are zero. See the right graph for the resulting model. This model can be collapsed over Age, as Age is not on any short path, compare Section \ref{collapse}. Here we implicitly assume that coverage in \textit{at least one} of the other sources is homogeneous w.r.t. age, and this is clearly reasonable for MOH. Although the census has some differential response by age, this is much less extreme than the partial population coverage of DIA, and we consider it reasonable to treat this source as homogeneous too. However, even if there is an important dependence of census on age we can make valid estimates as long as the interaction between Census and MOH is also included in the model, as then age is not on a short path and the model can be collapsed over it.

The number of unique individuals in the three linked registers is 4,378,377, the estimated number not in any register is 40,868, and this leads to an estimated population size of 4,419,245 (4,415,848 -- 4,422,929). The estimated number of non--M\=aori and M\=aori in the Census, DIA and MOH can be found in Table \ref{summary}, Panel B.

\linespread{1.0}

\linespread{2.0}

Parameter estimates for the maximal model are found in Table \ref{parest3}. With $n = 4,378,377$ it is not surprising that it is hard to find more parsimonious models that fit the data, except for the model where the three-factor interaction between $C, a$ and $b$ is set to zero. Using the methodology that we propose it is possible to fit more parsimonious models but in this particular instance there is not much to gain. For three registers the maximal model allows for pairwise dependence of registers, whereas with two registers conditional independence had to be assumed. For example, for the pairwise dependence of register $A$ and $B$ we use the parameter estimates for $A$:$B$ and $A$:$B$:$c$. Then for non--M\=aori ($c = 0$) the estimated conditional odds ratio between $A$ and $B$ is exp(0.332) = 1.4, and for M\=aori it is exp$(0.332 - 0.1596) = 1.2$, so inclusion in $A$ goes together with inclusion in $B$. These relations are much stronger for $A$ and $C$, and for $B$ and $C$. For example, for $b=0$ the estimated conditional odds ratio between $A$ and $C$ is exp(2.0042) = 7.4, and for $b=1$ the estimated conditional odds ratio between $A$ and $C$ is exp$(2.0042-0.8567) = 3.2$. Not surprisingly, there are strong relations between the ethnicity variables $a$, $b$ and $c$. For parameters $a$:$b$, $a$:$c$ and $b$:$c$ the  estimates are 5.472, 5.075 and 4.156. Interestingly, the estimated three factor interaction parameter $a$:$b$:$c$ is negative. This means that, for example, for the non-M\=aori in $c$ the estimated odds-ratio between ethnicity measures in $a$ and $b$ is exp(5.4720) = 237.9, whereas for M\=aori in $c$ the estimated odds-ratio between the ethnicity measures in $a$ and $b$ is only exp$(5.4720 - 2.1362) = 28.1$. Thus the negative three-factor interaction estimate $a$:$b$:$c$ shows that for non-M\=aori in one of the three registers the disagreement in the other two registers is smaller than for M\=aori (i.e., M\=aori are more likely to be inconsistent between the registers than non-M\=aori). Table \ref{tab_abc} illustrates this for counts marginalized over the registers $A, B$ and $C$: for example, for $c=0$ the disagreement is 64,566 + 30,087 (relative to approximately 3,750,000), but for $c=1$ the disagreement is 26,063 + 18,448 (relative to approximately 625,000).

\linespread{1.0}
\begin{table}[h!]
\bf \caption{\rm \it Parameter estimates for three registers model}
\label{parest3}\rm
\begin{center}
\begin{tabular}{lrrrr}
\hline
            &Estimate&Std. Error& z value& Pr\\
\hline
(Intercept) &  10.487     &NA        &NA     &NA\\
A          &   0.029 &0.050 &    0.6  &0.5592\\
B          &$ -4.001$&0.033 &$-122.4$ &$<0.001$\\
c          &$ -4.866$&0.025 &$-198.5$ &$<0.001$\\
C          &   2.254 &0.050 &   45.1  &$<0.001$\\
b          &$ -4.303$&0.183 & $-23.5$ &$<0.001$\\
a          &$ -5.756$&0.201 & $-28.7$ &$<0.001$\\
A:B        &   0.332 &0.005 &   67.9  &$<0.001$\\
A:c        &$ -0.750$&0.024 & $-31.7$ &$<0.001$\\
B:c        &   1.080 &0.014 &   79.4  &$<0.001$\\
A:C        &   2.004 &0.050 &   40.1  &$<0.001$\\
A:b        &   0.426 &0.086 &    4.9  &$<0.001$\\
C:b        &   0.661 &0.182 &    3.6  &$<0.001$\\
B:C        &   2.030 &0.032 &   62.7  &$<0.001$\\
B:a        &   1.700 &0.059 &   28.8  &$<0.001$\\
C:a        &   0.702 &0.201 &    3.5  &$<0.001$\\
c:b        &   4.156 &0.032 &  128.9  &$<0.001$\\
c:a        &   5.075 &0.022 &  232.1  &$<0.001$\\
b:a        &   5.472 &0.275 &   19.9  &$<0.001$\\
A:B:c      &$ -0.160$&0.008 & $-19.7$ &$<0.001$\\
A:C:b      &$ -0.857$&0.086 & $-10.0$ &$<0.001$\\
B:C:a      &$ -0.568$&0.059 &  $-9.6$ &$<0.001$\\
A:c:b      &   0.232 &0.029 &    8.1  &$<0.001$\\
B:c:a      &$ -1.129$&0.014 & $-81.7$ &$<0.001$\\
C:b:a      &$ -0.249$&0.275 &  $-0.9$ &0.3664\\
c:b:a      &$ -2.136$&0.029 & $-73.0$ &$<0.001$\\
\hline
\end{tabular}
\end{center}
\end{table}
\linespread{2.0}

\linespread{1.0}
\begin{table}[h!]
\bf \caption{\rm \it Contingency table of fitted frequencies for the marginal table of $a, b$ and $c$}
\label{tab_abc}\rm
\begin{center}
\begin{tabular}{l rr rr}
\hline
&\multicolumn{2}{c}{$c=0$}&\multicolumn{2}{c}{$c=1$}\\
&$b=0$&$b=1$&$b=0$&$b=1$\\
\hline
$a=0$              &   3,581,229  &  64,566  &  18,264  &  26,063 \\
$a=1$              &      30,087  & 100,639  &  18,448  & 579,949 \\
\hline
\end{tabular}
\end{center}
\end{table}
\linespread{2.0}

\subsection{Four registers}
\label{4reg}
We now add the fourth register, MOE, denoted $D$, with ethnicity in MOE denoted as $d$, to the analysis. Now the maximal model is $[ABCd][ABDc][ACDb]$
$[BCDa][ABcd] \allowbreak [ACbd][ADbc][BCad][BDac][CDab][Abcd][Bacd][Cabd][Dabc]$
$[abcd]$. Due to the appearance of four factor interactions in the model the assumptions become less and less demanding as more registers are involved.

In this model both DIA and MOE have heterogeneous inclusion probabilities over Age. We argue here that this heterogeneity does not threaten the validity of our estimates. Assuming homogeneous inclusion probabilities over age for the Census and MOH, the heterogeneity would lead to a model where Age is related to both DIA and MOE, but not the other registers. Therefore Age is not on a short path (comparable to Figure \ref{Fig4new}, model M17) and we can collapse over Age. We conclude that the fact that DIA and MOE do not aim to cover the full population, does not threaten the validity of our estimates.

This model turns out to have some numerical instability in the sense that some log-linear parameters are (nearly) on the boundary or non-identified. We restricted 15 parameters to zero leading to the model $[ABcd][AC][ADbc][BCad]$ $[BDac][CDa][CDb][Abcd][Bacd][Dabc]\allowbreak[abcd]$. This stabilizes the estimates. The deviance of the model is 680.6, which is, with an observed population size of over 4 million, negligible. The population size estimates for the model are in Table \ref{summary}, Panel C. The loglinear parameter estimates for the maximal model and for the restricted version are in the supplementary materials. The number of unique individuals in the four linked registers is 4,401,990, and the estimated number missed by all registers is 20,972, giving an estimated population size of 4,422,962 (4,421,894 -- 4,424,080). In comparison to the three register solution, by including the fourth register the estimated population size increases by only 4,000, mostly M\=aori.

\linespread{1.0}

\linespread{1.0}

\subsection{Three registers, ignoring the Census}
\label{nocensus}
We also present estimates derived only from the three administrative registers, so that we can see what would happen if the census were replaced entirely by an administrative data-based system. The observed number of individuals in at least one of the registers is 4,378,716. We estimate an additional 26,513 individuals missed by all three registers. This leads to a total population size of 4,405,229 (4,401,858 -- 4,409,667). This is somewhat less than the four register estimate, that was 4,422,962.

The estimate of 4,405,229 broken down by ethnicity for each of the three registers is presented in Table \ref{summary}, Panel D. The three-register M\=aori estimate for DIA is larger than the four-register estimate in Table \ref{summary}, Panel C: 804,936 versus 775,697; for MOH it is very similar and for MOE the three-register estimate is larger as well: 780,234 versus 762,222. This suggests that in the absence of the census, the estimate of the size of the M\=aori population would be larger. Based on the estimates of measurement error from the latent variable analysis in section \ref{measurement error}, where the census was the most accurate, this suggests an overestimation of the M\=aori population size.

\linespread{1.0}

\linespread{2.0}

\section{Dealing with measurement error in M\=aori variables}
\label{measurement error}

To arrive at a final estimate of the number of non--M\=aori and M\=aori we describe two approaches, both using the concept of measurement error. Deriving a final estimate is related to the field of macro--integration of categorical outcomes, see de Waal et al. (2019 and in press) for a review of the field. Consider the joint margins of the ethnicity variables $a, b, c$ and $d$ of the four registers in Table \ref{margabcd} from the restricted maximal model of Section \ref{4reg}. An ad hoc approach is to consider five groups of individuals, according to how many registers record them as M\=aori. Then a (slightly arbitrary) choice would have to be made which categories would be used to arrive at the estimated numbers of M\=aori and non--M\=aori. For example, one could consider individuals who are estimated to be M\=aori in at least two registers to be M\=aori, allowing for a measurement error in recording a non--M\=aori as a M\=aori in at most one register. This would give a final estimate of 768,479, corresponding to an estimated probability of $768,479/4,422,962 = 0.174$.

\linespread{1.0}
\begin{table}
\bf \caption{\rm\textit{Joint margins for variables $a$ (Census), $b$ (DIA), $c$ (MOH) and $d$ (MOE) estimated under the restricted maximal model}.}
\label{margabcd}\rm
\begin{center}
\begin{tabular}{ll rr rr}
\\ \hline
      &      &\multicolumn{2}{c}{$c=0$}&\multicolumn{2}{c}{$c=1$}\\
      &      &$d=0$         &$d=1$      &$d=0$     &$d=1$  \\
\hline\\
$a=0$  &$b=0$ & 3,519,852    &    53,366 &  10,998  & 6,934\\
      &$b=1$ &    55,676    &    15,600 &   9,686  & 17,555 \\
\\
$a=1$  & $b=0$&    14,590    &    21,218 &   2,560  & 17,747 \\
      & $b=1$&    18,443    &    79,105 &  28,934  &550,697 \\
\hline
\end{tabular}
\end{center}
\end{table}
\linespread{2.0}

A statistical approach to measurement error is to make use of a latent class model (McCutcheon, 1987)\nocite{McCutcheon1987}, a technique that has also been applied to evaluate census-related multiple system estimation by Biemer et al. (2001)\nocite{BiemerWoltmannRaglinHill2001}, see for comparable work Biggeri et al. (1999)\nocite{BiggeriStanghelliniMerlettiMarchi1999}, Stanghellini and van der Heijden (2004)\nocite{StanghellinivanderHeijden2004} and di Cecco et al. (2018)\nocite{diCeccodiZioFilipponiRoccheti2018}. See also Boeschoten, de Waal and Vermunt (2019)\nocite{BoeschotendeWaalVermunt2019}, for a comparison of the latent class model with the ad hoc approach described in the preceding paragraph. Other recent work making use of the latent class model in official statistics is Boeschoten, Oberski and de Waal (2017)\nocite{BoeschotenOberskideWaal2017} and Boeschoten, de Waal and Vermunt (2019)\nocite{BoeschotendeWaalVermunt2019}. Oberski (2018)\nocite{Oberski} makes a plea for the use of latent class modelling in the context of linked data sources to tackle the measurement error problem.

The latent class model assumes the existence of a categorical latent variable, and that the observed variables are independent conditional on the latent variable. Thus the latent variable ``causes'' the responses to the observed variables, and explains the interactions between the observed variables. Let $\pi_{abcd}$ be the joint probability for variables $a, b, c$ and $d$, with levels indexed by $r, s, t$ and $u$ respectively, with 0 for non--M\=aori and 1 for M\=aori. Let $X$ be the latent variable, with two levels indexed by $x$ ($x=1,2$). Let $\pi^X_{x}$ be the probability to fall in latent class $x$. Let $\pi^a_{r|x}$ be the conditional probability for variable $a$ to fall in $r = 0,1$ given latent class $x$. Then the two-class latent class model is
\begin{equation}
\label{LCAmodel}
\pi_{rstu} = \Sigma_{x=1,2} \pi^X_{x}\pi^{a}_{r|x}\pi^{b}_{s|x}\pi^{c}_{t|x}\pi^{d}_{u|x}.
\end{equation}

As there are 16 counts in Table \ref{margabcd}, and nine parameters in equation \ref{LCAmodel} (i.e. eight independent conditional probabilities and one independent class size parameter), there are 7 degrees of freedom.

We fit the latent class model with two latent classes to the data in Table \ref{margabcd}. This gives a two-stage procedure - first fitting a model to deal with the missing data (both $-$ and x), and then applying a second, latent class, model to the resulting estimates. We find the estimates in Panel 1 of Table \ref{lcaest}. In this latent class model, the first latent class is to be interpreted as the class for non--M\=aori, and the estimated probability of falling in this class is 0.827. The estimated probability for the M\=aori class, 0.1733, corresponds to an estimated M\=aori population size of approximately 770,677. Estimated conditional probabilities of being M\=aori for each latent class are also shown in Panel 1 of Table \ref{lcaest}; they are consistently low for the non--M\=aori latent class and high for the M\=aori latent class. These estimated conditional probabilities can be interpreted as measurement error estimates under the assumption that the model is correct. For example, with $1 - 0.937 = 0.063$ the census has the smallest measurement error for M\=aori: given that the true status of someone is M\=aori, he or she has an estimated probability of 0.063 to say he or she is non--M\=aori in the census. For the non--M\=aori the MOH has the smallest measurement error, with an estimate of 0.003, closely followed by the census. The measurement errors are much larger in the latent class of the M\=aori than in the class of the non--M\=aori. The population size of the M\=aori latent class of 770,677 may be considered high when we compare it with the estimates for the Census, DIA, MOH and MOE of 733,167, 761,545, 643,429 and 770,047 respectively. However, we have to take into account that in this M\=aori latent class individuals have relatively large measurement error, making them report regularly that they are non-M\=aori. The reverse measurement error, that individuals in the non-M\=aori latent class report that they are M\=aori, is much smaller.  We note that a three class latent class model is not identified if the number of observed variables is four, even though the number of cells is larger than the number of parameters (Vermunt and Magidson, 2004)\nocite{VermuntMagidson2004}.

\linespread{1.0}
\begin{table}
\bf \caption{\rm\textit{Estimates of latent class models with two latent classes}.}
\label{lcaest}\rm
\begin{center}
\begin{tabular}{lrrrrr}
\multicolumn{6}{l}{\it Panel 1: Estimates for Table \ref{margabcd}}\\
\\
\hline
        &          &census            &DIA                &MOH                &MOE     \\
        &$\pi_{x}$ &$\pi_{r=1|x}^{a}$ & $\pi_{s=1|x}^{b}$ & $\pi_{t=1|x}^{c}$ & $\pi_{u=1|x}^{d}$ \\
\hline
\\
Class 1 &0.827              & 0.004        &0.016      &0.003      &0.015 \\
Class 2 &0.173              & 0.937        &0.937      &0.826      &0.922 \\
\hline
\\
\multicolumn{6}{l}{\it Panel 2: Estimates for LCMSE}\\
\hline
\\
Class 1 &0.834              & 0.007        &0.014      &0.005      &0.016 \\
Class 2 &0.166              & 0.957        &0.958      &0.847      &0.959 \\
\hline
\end{tabular}
\end{center}
\end{table}
\linespread{2.0}

We have just fitted a latent class model with two classes on the margins for variables $a, b, c$ and $d$. It is also possible to define overall models. For this purpose we first rewrite the latent class model with two latent classes as a loglinear model for the observed variables $a, b, c$ and $d$ and the latent variable $X$ (Hagenaars, 1993)\nocite{HagenaarsJA1993}. The latent class model can be denoted as the loglinear model $[aX][bX][cX][dX]$ for the probabilities $\pi_{rstux}$, where $\Sigma_x \pi_{rstux} = \pi_{rstu}$. Written as a loglinear model using $\lambda$--parameters, we have
\begin{equation}
\label{LLLCA-model}
\log \pi_{rstux} = \lambda + \lambda^a_r + \lambda^b_s + \lambda^c_t + \lambda^d_u + \lambda^X_x + \lambda^{aX}_{rx} + \lambda^{bX}_{sx} + \lambda^{cX}_{tx} + \lambda^{dX}_{ux}
\end{equation}

with
\begin{equation}
\label{sumLL-model}
\pi_{rstu} = \Sigma_x \pi_{rstux}.
\end{equation}

We incorporate this loglinear model in the maximal model for variables $A, B, C, D, a, b, c$ and $d$ in the following way. To start with, we define probabilities $\pi_{rstuRSTUx}$ that allow us to define a model for the eight observed variables and the latent variable. These probabilities that include a latent variable are related to the probabilities for the observed variables only by
\begin{equation}
\label{sumLL2-model}
\pi_{rstuRSTU} = \Sigma_x \pi_{rstuRSTUx}.
\end{equation}

In the section on four registers we saw that the maximal model is $[ABCd]$ $[ABDc] \allowbreak[ACDb][BCDa][ABcd][ACbd][ADbc][BCad][BDac][CDab][Abcd]$ $[Bacd][Cabd][Dabc][abcd]$. The restrictive features of this maximal model are (1) that variables denoted by a lower case letter cannot appear in the same term as variables denoted by capitals, so that, for example, variables $a$ and $A$ cannot appear in the same interaction term, and (2) the four capitals $A, B, C$ and $D$ cannot appear together in the same term. We are particularly interested in the joint marginal counts of $a, b, c$ and $d$.

We now modify the maximal model to include a latent variable. In the maximal model there are three groups of terms. First, there is the last term $[abcd]$, that we replace by the latent class model $[aX][bX][cX][dX]$. Thus we are formulating a model for observed variables and the latent variable. Second, we eliminate the terms $[ABcd], [ACbd], [ADbc], [BCad], \allowbreak [BDac], [CDab], [Abcd], [Bacd]$, $[Cabd], [Dabc]$ as the joint appearance of two or three of the lower case letters for variables $a, b, c$ and $d$ in one single term is in conflict with the latent class assumption. For example, consider $[ABcd]$. The interaction between $c$ and $d$ is explained by the latent variable $X$. One could argue that, although it is in conflict with the latent class model to include the $c,d-$interaction, one could still include the interactions for $A,c,d$, for $B,c,d$ and for $A,B,c,d$. However, this would result in a non--hierarchical loglinear model and these are difficult to interpret, so we eliminate  these terms completely. Third, we can retain the terms $[ABCd], [ABDc], [ACDb], [BCDa]$. Thus a latent class model where (\ref{LLLCA-model}) is extended with the register variables $A, B, C$ and $D$, is $[ABCd][ABDc][ACDb][BCDa][aX][bX][cX][dX]$. We refer to this model as LCMSE, short for latent class multiple system estimation.

Our code for the LCMSE model can be found in the supplementary materials B and C. The LCMSE model was also fitted using lEM (Vermunt, 2007), however, lEM does not provide an easy way to calculate estimates of the missing part of the population.

The latent class parameter estimates can be found in Panel 2 of Table \ref{lcaest}. They are similar to those in Panel 1. The deviance of the model is 10,922.25. A proper evaluation of this value should take the observed population size of 4,401,990 into account, and therefore we divide the deviance by 4,401.99, so that we have an impression of the deviance if the population size had been 1,000. Thus we find a normed deviance of 2.5, showing a good fit. The population size estimate for this model population is 4,447,071, with a 95 percent confidence interval of 4,435,301 -- 4,465,050. The estimated proportion of Maori in the LCMSE is 0.166, lower than in the two-stage model, and the non--M\=aori proportion is correspondingly higher. It is as if the LCMSE approach is more likely to treat individuals with uncertain status as part of the non--M\=aori latent class. This reduces the estimated measurement errors in the M\=aori latent class, and increases them in the non-M\=aori. We prefer the integrated approach of the LCMSE because it deals with all the variability in the data appropriately, whereas the two-step approach means that the latent class estimates depend on the fitted values of the first stage model.


As a second model, we adjust the LCMSE model by including a latent variable $Y$ that explains the relations between the variables $A, B, C$ and $D$. Theoretically, such a model allows for heterogeneity of inclusion probabilities, where in one class inclusion probabilities for the four registers are higher and in the other class they are lower, see Stanghellini and van der Heijden (2004)\nocite{StanghellinivanderHeijden2004} for a medical example. So if there were two such subpopulations, the latent class model would reveal this. A model for this is $[AY][BY][CY][DY][aX][bX][cX][dX]$ (now excluding all of the original interactions as these are explained by the latent class variables). This model has a deviance of 291,464.1, and a normed deviance of 66.2. Thus we reject this model. We also considered extending the model with an interaction between $X$ and $Y$, giving $[AY][BY][CY][DY][aX][bX][cX][dX][XY]$. The fit improves but is still not good: the normed deviance is 43.7.

We conclude that the LCMSE model is the best model for these data. Interestingly, the estimated probability of the M\=aori latent class of 0.166 in this model is close to the estimated probability of M\=aori of 16.5 for the IDI-ERP found by Statistics New Zealand. Also, the tables analysed are for Census day 5th March 2013. The official ERP figure for March 2013 is 4,436,000 +/- about 0.5 percent  (https://www.stats.govt.nz/topics/population). This is very close to our 4 register estimate of 4,422,962 and the LCMSE estimate of 4,447,071.

\section{Discussion}
\label{discussion}
Van der Heijden et al. (2018) presented an approach for estimating the margins of auxiliary variables in the dual system estimation framework. They suggested that more experience with applications of this methodology was needed to be able to judge its usefulness. Here this approach is extended to multiple system estimation with four registers, and a more complicated missing data structure.

The data used here were probabilistically linked in Stats NZ’s Integrated Data Infrastructure (IDI), which means that the linkage has not been subjected to a clerical review stage. This increases the risk that there will be some linkage error which is not accounted for in our models, and since dual and multiple system estimation using loglinear models is critically dependent on the lack of linkage error (Wolter, 1986\nocite{WolterKM1986}, Biemer and Stokes, 2004 \nocite{BiemerStokes2004}, Gerritse et al., 2017\nocite{GerritseBakkerHeijden2017}), there is a risk that estimates will be inflated through matching error. The data have also been randomly rounded to base 3 to protect confidentiality, but we expect this to have a negligible impact on the fitted values as the number of individuals in the linked data set is larger than 4 million.

There is also a risk of overcoverage (also known as list inflation) in administrative registers, through duplicate records and the inclusion of people who are not part of the target population, which is the usual resident population in the case of the census as considered here.  It is likely that some duplicates remain in the IDI-ERP despite attempts to remove as many of them as possible. There is as yet no estimate of the remaining overcoverage in the IDI-ERP. However, among the sources considered here the DIA birth register is less likely to suffer from duplicates because it does not obtain repeating data -- though for the same reason it may suffer from overcoverage when people emigrate. However, the sequence of population size estimates from models with two to four registers shows only small increases, suggesting that the overcoverage is small in the added registers.

We assume that the inclusion probability is homogeneous in at least some of the registers (see section \ref{3reg}). This assumption seems most at risk for the census source with respect to age, though as we discussed previously this does not invalidate our analysis because we have homogeneity in other sources; at worst it may restrict our choice of models to ensure that age does not appear on a short path between registers. However, even in the case that age was important in \textit{all} the registers, it would still be possible to use missing data methodology to estimate the unobserved parts of the registers under the assumption that the relations in parts of the population where the registers overlap also hold in the other parts - a missing at random assumption (Zwane et al. 2004).

The modelling process to choose the most parsimonious loglinear model to use in population size estimation and marginal total estimation for auxiliary variables such as M\=aori/non--M\=aori is complicated in our application by the size of the data. For most of the models except for the model for four sources, since the census and the MOH registers cover most of the population of New Zealand, the entries in the contingency tables (for example Table \ref{missing}, Panel 1) are very large. Therefore any term added to the model has plenty of data for estimation, and will almost certainly be significant. Therefore we have ended up with saturated models, except for the model for four sources where the saturated model is numerically unstable, and the latent class model, that is not saturated by definition. The estimated sizes of the M\=aori population are plausible, but a different value is created according to the definition in each register (in this sense the M\=aori identifiers act as alternative variables, as in van der Heijden et al. (2018, section 4)). A preferred version of the variable, or some pragmatic combination of estimates must be chosen in order to obtain a final estimate.

The latent class analysis of the same data deals with these different definitions, and provides a consolidated single estimate, and, as a consequence, also estimates of the measurement errors in each source. In the two models for which estimates are presented the measurement errors for M\=aori status in the census are among the lowest (Table \ref{lcaest}, Panel 1 and 2), which provides some support for the intuition in the statistical system that the purpose-designed census collection gives a better estimate of the size of the M\=aori population than one based on administrative registers which only collect this variable as a by-product of their main purpose. The preferred LCMSE model (Table \ref{lcaest}, Panel 2) shows almost the same estimated measurement error for the M\=aori in census, DIA and MOE, with MOH showing a substantially greater measurement error. MOH has the lowest estimated measurement  error for non-M\=aori, but only a little better than the Census (under either latent class approach).

Despite these differences, the estimate using only the three administrative registers without the input of the census is reasonable, although there are some differences from methods based on the census.

We conclude that the methods of van der Heijden et al. (2018) provide stable results that allow for detailed interpretation of the processes of inclusion in the registers considered, and of recording M\=aori status. The conditions for treating the partial coverage of the registers as ignorable are met in this study, so that the modelling can be applied directly, and we show that they can be extended to deal with different forms of missing data. The latent class approach provides a principled method to produce a common estimate accounting for differences in the definition of M\=aori status among the data sources, and also provides estimates of the measurement error in the different definitions which can be used to understand the quality of the administrative sources.

\section{Disclaimer}

The results in this paper are not official statistics, they have been created for research purposes from the Integrated Data Infrastructure (IDI) managed by Stats NZ. The opinions, findings, recommendations and conclusions expressed in this paper are those of the authors, not Stats NZ. Access to the anonymised data used in this study was provided by Stats NZ in accordance with security and confidentiality provisions of the Statistics Act 1975. Only people authorised by the Statistics Act 1975 are allowed to see data about a particular person, household, business or organisation and the results in this paper have been confidentialised to protect these groups from identification. Careful consideration has been given to the privacy, security and confidentiality issues associated with using administrative and survey data in the IDI. Further detail can be found in the Privacy impact assessment for the Integrated Data Infrastructure (Statistics New Zealand, 2017b)\nocite{StatisticsNewZealand2017b}.

\section{Supplementary Materials}
\begin{description}
\item[Supplementary Material A:]
Raw data tables and parameter estimates. (file type: .pdf)
\item[Supplementary Material B:]
R Package (including data): cenzus. Openly available at \url{https://github.com/MaartenCruyff/cenzus}.
\item[Supplementary Material C:]
Computer code and extended output of parameter estimates (file type: .html)
\end{description}

\clearpage

\bibliographystyle{apacite}
\setstretch{1.0}
\bibliography{BibPeterUtrecht1.bib}

\providecommand{\noopsort}[1]{}
\begin{thebibliography}{}

\bibitem [\protect \citeauthoryear {%
Bakker%
, van~der Heijden%
\BCBL {}\ \BBA {} Scholtus%
}{%
Bakker%
\ \protect \BOthers {.}}{%
{\protect \APACyear {2015}}%
}]{%
BakkerBFMvanderHeijdenPGMScholtusS2015}
\APACinsertmetastar {%
BakkerBFMvanderHeijdenPGMScholtusS2015}%
\begin{APACrefauthors}%
Bakker, B.%
, van~der Heijden, P.%
\BCBL {}\ \BBA {} Scholtus, S.%
\end{APACrefauthors}%
\unskip\
\newblock
\APACrefYearMonthDay{2015}{}{}.
\newblock
{\BBOQ}\APACrefatitle {Preface to Special Issue on Coverage Problems in
  Administrative Sources} {Preface to special issue on coverage problems in
  administrative sources}.{\BBCQ}
\newblock
\APACjournalVolNumPages{Journal of Official Statistics}{31}{}{349-355}.
\PrintBackRefs{\CurrentBib}

\bibitem [\protect \citeauthoryear {%
P.~Biemer%
, Woltman%
, Raglin%
\BCBL {}\ \BBA {} Hill%
}{%
P.~Biemer%
\ \protect \BOthers {.}}{%
{\protect \APACyear {2001}}%
}]{%
BiemerWoltmannRaglinHill2001}
\APACinsertmetastar {%
BiemerWoltmannRaglinHill2001}%
\begin{APACrefauthors}%
Biemer, P.%
, Woltman, H.%
, Raglin, D.%
\BCBL {}\ \BBA {} Hill, J.%
\end{APACrefauthors}%
\unskip\
\newblock
\APACrefYearMonthDay{2001}{}{}.
\newblock
{\BBOQ}\APACrefatitle {Enumeration accuracy in a population census: An
  evaluation using latent class analysis} {Enumeration accuracy in a population
  census: An evaluation using latent class analysis}.{\BBCQ}
\newblock
\APACjournalVolNumPages{Journal of Official Statistics}{17}{}{129--148}.
\PrintBackRefs{\CurrentBib}

\bibitem [\protect \citeauthoryear {%
P\BPBI P.~Biemer%
\ \BBA {} Stokes%
}{%
P\BPBI P.~Biemer%
\ \BBA {} Stokes%
}{%
{\protect \APACyear {2004}}%
}]{%
BiemerStokes2004}
\APACinsertmetastar {%
BiemerStokes2004}%
\begin{APACrefauthors}%
Biemer, P\BPBI P.%
\BCBT {}\ \BBA {} Stokes, S\BPBI L.%
\end{APACrefauthors}%
\unskip\
\newblock
\APACrefYearMonthDay{2004}{}{}.
\newblock
{\BBOQ}\APACrefatitle {Approaches to the modeling of measurement error}
  {Approaches to the modeling of measurement error}.{\BBCQ}
\newblock
\BIn{} P\BPBI P.~Biemer, R\BPBI M.~Groves, L\BPBI E.~Lyberg, N\BPBI
  A.~Mathiowetz\BCBL {}\ \BBA {} S.~Sudman\ (\BEDS), \APACrefbtitle
  {Measurement error in surveys.} {Measurement error in surveys.}
\newblock
\APACaddressPublisher{New York}{Wiley}.
\PrintBackRefs{\CurrentBib}

\bibitem [\protect \citeauthoryear {%
Biggeri%
, Stanghellini%
, Merletti%
\BCBL {}\ \BBA {} Marchi%
}{%
Biggeri%
\ \protect \BOthers {.}}{%
{\protect \APACyear {1999}}%
}]{%
BiggeriStanghelliniMerlettiMarchi1999}
\APACinsertmetastar {%
BiggeriStanghelliniMerlettiMarchi1999}%
\begin{APACrefauthors}%
Biggeri, A.%
, Stanghellini, E.%
, Merletti, F.%
\BCBL {}\ \BBA {} Marchi, M.%
\end{APACrefauthors}%
\unskip\
\newblock
\APACrefYearMonthDay{1999}{}{}.
\newblock
{\BBOQ}\APACrefatitle {Latent class models for varying catchability and
  correlation among sources in capture--recapture estimation of the size of a
  human population.} {Latent class models for varying catchability and
  correlation among sources in capture--recapture estimation of the size of a
  human population.}{\BBCQ}
\newblock
\APACjournalVolNumPages{Statistica Applicata}{11}{}{563--576}.
\PrintBackRefs{\CurrentBib}

\bibitem [\protect \citeauthoryear {%
Boeschoten%
, {de Waal}%
\BCBL {}\ \BBA {} Vermunt%
}{%
Boeschoten%
\ \protect \BOthers {.}}{%
{\protect \APACyear {2019}}%
}]{%
BoeschotendeWaalVermunt2019}
\APACinsertmetastar {%
BoeschotendeWaalVermunt2019}%
\begin{APACrefauthors}%
Boeschoten, L.%
, {de Waal}, T.%
\BCBL {}\ \BBA {} Vermunt, J\BPBI K.%
\end{APACrefauthors}%
\unskip\
\newblock
\APACrefYearMonthDay{2019}{}{}.
\newblock
{\BBOQ}\APACrefatitle {Estimating the number of serious road injuries per
  vehicle type in the {N}etherlands by using multiple imputation of latent
  classes} {Estimating the number of serious road injuries per vehicle type in
  the {N}etherlands by using multiple imputation of latent classes}.{\BBCQ}
\newblock
\APACjournalVolNumPages{Journal of the Royal Statistal Society, Series
  A}{182}{}{1463--1486}.
\PrintBackRefs{\CurrentBib}

\bibitem [\protect \citeauthoryear {%
Boeschoten%
, Oberski%
\BCBL {}\ \BBA {} {de Waal}%
}{%
Boeschoten%
\ \protect \BOthers {.}}{%
{\protect \APACyear {2017}}%
}]{%
BoeschotenOberskideWaal2017}
\APACinsertmetastar {%
BoeschotenOberskideWaal2017}%
\begin{APACrefauthors}%
Boeschoten, L.%
, Oberski, D.%
\BCBL {}\ \BBA {} {de Waal}, T.%
\end{APACrefauthors}%
\unskip\
\newblock
\APACrefYearMonthDay{2017}{}{}.
\newblock
{\BBOQ}\APACrefatitle {Estimating Classification Errors Under Edit Restrictions
  in Composite Survey-Register Data Using Multiple Imputation Latent Class
  Modelling {(MILC)}} {Estimating classification errors under edit restrictions
  in composite survey-register data using multiple imputation latent class
  modelling {(MILC)}}.{\BBCQ}
\newblock
\APACjournalVolNumPages{Journal of Official Statistics}{33}{}{921--962}.
\PrintBackRefs{\CurrentBib}

\bibitem [\protect \citeauthoryear {%
Brown%
, Diamond%
, Chambers%
, Buckner%
\BCBL {}\ \BBA {} Teague%
}{%
Brown%
\ \protect \BOthers {.}}{%
{\protect \APACyear {1999}}%
}]{%
BrownJJDiamondIDChambersRL1999}
\APACinsertmetastar {%
BrownJJDiamondIDChambersRL1999}%
\begin{APACrefauthors}%
Brown, J.%
, Diamond, I.%
, Chambers, R.%
, Buckner, L.%
\BCBL {}\ \BBA {} Teague, A.%
\end{APACrefauthors}%
\unskip\
\newblock
\APACrefYearMonthDay{1999}{}{}.
\newblock
{\BBOQ}\APACrefatitle {A methodological strategy for a one-number census in the
  {UK}} {A methodological strategy for a one-number census in the {UK}}.{\BBCQ}
\newblock
\APACjournalVolNumPages{Journal of the Royal Statistical Society: Series A
  (Statistics in Society)}{162}{2}{247--267}.
\newblock
\begin{APACrefDOI} \doi{10.1111/1467-985X.00133} \end{APACrefDOI}
\PrintBackRefs{\CurrentBib}

\bibitem [\protect \citeauthoryear {%
Brown%
, Sexton%
, Abbott%
\BCBL {}\ \BBA {} Smith%
}{%
Brown%
\ \protect \BOthers {.}}{%
{\protect \APACyear {2019}}%
}]{%
BrownJSextonCAbbottOSmithPA2019}
\APACinsertmetastar {%
BrownJSextonCAbbottOSmithPA2019}%
\begin{APACrefauthors}%
Brown, J.%
, Sexton, C.%
, Abbott, O.%
\BCBL {}\ \BBA {} Smith, P\BPBI A.%
\end{APACrefauthors}%
\unskip\
\newblock
\APACrefYearMonthDay{2019}{}{}.
\newblock
{\BBOQ}\APACrefatitle {The framework for estimating coverage in the 2011
  {C}ensus of {E}ngland and {W}ales: combining dual-system estimation with
  ratio estimation} {The framework for estimating coverage in the 2011 {C}ensus
  of {E}ngland and {W}ales: combining dual-system estimation with ratio
  estimation}.{\BBCQ}
\newblock
\APACjournalVolNumPages{Statistical Journal of the IAOS}{35}{}{481--499}.
\newblock
\begin{APACrefDOI} \doi{10.3233/SJI-180426} \end{APACrefDOI}
\PrintBackRefs{\CurrentBib}

\bibitem [\protect \citeauthoryear {%
Buckland%
\ \BBA {} Garthwaite%
}{%
Buckland%
\ \BBA {} Garthwaite%
}{%
{\protect \APACyear {1991}}%
}]{%
BucklandSGarthwaiteP1991}
\APACinsertmetastar {%
BucklandSGarthwaiteP1991}%
\begin{APACrefauthors}%
Buckland, S.%
\BCBT {}\ \BBA {} Garthwaite, P.%
\end{APACrefauthors}%
\unskip\
\newblock
\APACrefYearMonthDay{1991}{}{}.
\newblock
{\BBOQ}\APACrefatitle {Quantifying precision of mark-recapture estimates using
  the bootstrap and related methods} {Quantifying precision of mark-recapture
  estimates using the bootstrap and related methods}.{\BBCQ}
\newblock
\APACjournalVolNumPages{Biometrics}{47}{}{255-268}.
\newblock
\begin{APACrefDOI} \doi{10.2307/2532510} \end{APACrefDOI}
\PrintBackRefs{\CurrentBib}

\bibitem [\protect \citeauthoryear {%
de Waal%
, van Delden%
\BCBL {}\ \BBA {} Scholtus%
}{%
de Waal%
\ \protect \BOthers {.}}{%
{\protect \APACyear {2019}}%
}]{%
deWaalvanDeldenScholtus2019}
\APACinsertmetastar {%
deWaalvanDeldenScholtus2019}%
\begin{APACrefauthors}%
de Waal, T.%
, van Delden, A.%
\BCBL {}\ \BBA {} Scholtus, S.%
\end{APACrefauthors}%
\unskip\
\newblock
\APACrefYearMonthDay{2019}{}{}.
\newblock
{\BBOQ}\APACrefatitle {Quality measures for multisource statistics} {Quality
  measures for multisource statistics}.{\BBCQ}
\newblock
\APACjournalVolNumPages{Statistical Journal of the IAOS}{35}{}{179--192}.
\PrintBackRefs{\CurrentBib}

\bibitem [\protect \citeauthoryear {%
de Waal%
, van Delden%
\BCBL {}\ \BBA {} Scholtus%
}{%
de Waal%
\ \protect \BOthers {.}}{%
{\protect \APACyear {2020}}%
}]{%
deWaalvanDeldenScholtus2020}
\APACinsertmetastar {%
deWaalvanDeldenScholtus2020}%
\begin{APACrefauthors}%
de Waal, T.%
, van Delden, A.%
\BCBL {}\ \BBA {} Scholtus, S.%
\end{APACrefauthors}%
\unskip\
\newblock
\APACrefYearMonthDay{2020}{}{}.
\newblock
{\BBOQ}\APACrefatitle {Multi-source statistics: basic situations and methods}
  {Multi-source statistics: basic situations and methods}.{\BBCQ}
\newblock
\APACjournalVolNumPages{International Statistical Review}{88}{}{203--228}.
\PrintBackRefs{\CurrentBib}

\bibitem [\protect \citeauthoryear {%
di Cecco%
, di Zio%
, Filipponi%
\BCBL {}\ \BBA {} Rocchetti%
}{%
di Cecco%
\ \protect \BOthers {.}}{%
{\protect \APACyear {2018}}%
}]{%
diCeccodiZioFilipponiRoccheti2018}
\APACinsertmetastar {%
diCeccodiZioFilipponiRoccheti2018}%
\begin{APACrefauthors}%
di Cecco, D.%
, di Zio, M.%
, Filipponi, D.%
\BCBL {}\ \BBA {} Rocchetti, I.%
\end{APACrefauthors}%
\unskip\
\newblock
\APACrefYearMonthDay{2018}{}{}.
\newblock
{\BBOQ}\APACrefatitle {Population Size Estimation Using Multiple Incomplete
  Lists with Overcoverage} {Population size estimation using multiple
  incomplete lists with overcoverage}.{\BBCQ}
\newblock
\APACjournalVolNumPages{Journal of Official Statistics}{34}{2}{557 - 572}.
\PrintBackRefs{\CurrentBib}

\bibitem [\protect \citeauthoryear {%
di Cecco%
, di Zio%
\BCBL {}\ \BBA {} Liseo%
}{%
di Cecco%
\ \protect \BOthers {.}}{%
{\protect \APACyear {2020}}%
}]{%
diCeccodiZioLiseo2020}
\APACinsertmetastar {%
diCeccodiZioLiseo2020}%
\begin{APACrefauthors}%
di Cecco, D.%
, di Zio, M.%
\BCBL {}\ \BBA {} Liseo, B.%
\end{APACrefauthors}%
\unskip\
\newblock
\APACrefYearMonthDay{2020}{}{}.
\newblock
{\BBOQ}\APACrefatitle {Bayesian latent class models for capture–recapture in
  the presence of missing data} {Bayesian latent class models for
  capture–recapture in the presence of missing data}.{\BBCQ}
\newblock
\APACjournalVolNumPages{Biometrical Journal}{}{}{1 -- 13}.
\PrintBackRefs{\CurrentBib}

\bibitem [\protect \citeauthoryear {%
Gerritse%
, Bakker%
\BCBL {}\ \BBA {} {van der Heijden}%
}{%
Gerritse%
\ \protect \BOthers {.}}{%
{\protect \APACyear {2017}}%
}]{%
GerritseBakkerHeijden2017}
\APACinsertmetastar {%
GerritseBakkerHeijden2017}%
\begin{APACrefauthors}%
Gerritse, S\BPBI C.%
, Bakker, B\BPBI F\BPBI M.%
\BCBL {}\ \BBA {} {van der Heijden}, P\BPBI G\BPBI M.%
\end{APACrefauthors}%
\unskip\
\newblock
\APACrefYear{2017}.
\newblock
\APACrefbtitle {The impact of linkage errors and erroneous captures on the
  population size estimator due to implied coverage.} {The impact of linkage
  errors and erroneous captures on the population size estimator due to implied
  coverage.}
\newblock
\APACaddressPublisher{Den Haag}{Statistics Netherlands, discussion paper
  2017-16}.
\PrintBackRefs{\CurrentBib}

\bibitem [\protect \citeauthoryear {%
Hagenaars%
}{%
Hagenaars%
}{%
{\protect \APACyear {1993}}%
}]{%
HagenaarsJA1993}
\APACinsertmetastar {%
HagenaarsJA1993}%
\begin{APACrefauthors}%
Hagenaars, J\BPBI A.%
\end{APACrefauthors}%
\unskip\
\newblock
\APACrefYear{1993}.
\newblock
\APACrefbtitle {Loglinear Models with Latent Variables} {Loglinear models with
  latent variables}.
\newblock
\APACaddressPublisher{Newbury Park}{Sage}.
\PrintBackRefs{\CurrentBib}

\bibitem [\protect \citeauthoryear {%
Hand%
}{%
Hand%
}{%
{\protect \APACyear {2018}}%
}]{%
Hand2018}
\APACinsertmetastar {%
Hand2018}%
\begin{APACrefauthors}%
Hand, D\BPBI J.%
\end{APACrefauthors}%
\unskip\
\newblock
\APACrefYearMonthDay{2018}{}{}.
\newblock
{\BBOQ}\APACrefatitle {Statistical challenges of administrative and transaction
  data (with discussion)} {Statistical challenges of administrative and
  transaction data (with discussion)}.{\BBCQ}
\newblock
\APACjournalVolNumPages{Journal of the Royal Statistical Society, Series
  A}{181}{}{555-–605}.
\newblock
\begin{APACrefDOI} \doi{10.1111/rssa.12315} \end{APACrefDOI}
\PrintBackRefs{\CurrentBib}

\bibitem [\protect \citeauthoryear {%
Madden%
, Coleman%
, Mashford-Pringle%
\BCBL {}\ \BBA {} Connolly%
}{%
Madden%
\ \protect \BOthers {.}}{%
{\protect \APACyear {2019}}%
}]{%
MaddenRColemanCMashfordAConnolyM2019}
\APACinsertmetastar {%
MaddenRColemanCMashfordAConnolyM2019}%
\begin{APACrefauthors}%
Madden, R.%
, Coleman, C.%
, Mashford-Pringle, A.%
\BCBL {}\ \BBA {} Connolly, M.%
\end{APACrefauthors}%
\unskip\
\newblock
\APACrefYearMonthDay{2019}{}{}.
\newblock
{\BBOQ}\APACrefatitle {Indigenous identification: Past, present and a possible
  future} {Indigenous identification: Past, present and a possible
  future}.{\BBCQ}
\newblock
\APACjournalVolNumPages{Statistical Journal of the IAOS}{35}{}{23--27}.
\PrintBackRefs{\CurrentBib}

\bibitem [\protect \citeauthoryear {%
McCutcheon%
}{%
McCutcheon%
}{%
{\protect \APACyear {1987}}%
}]{%
McCutcheon1987}
\APACinsertmetastar {%
McCutcheon1987}%
\begin{APACrefauthors}%
McCutcheon, A.%
\end{APACrefauthors}%
\unskip\
\newblock
\APACrefYear{1987}.
\newblock
\APACrefbtitle {Latent class analysis} {Latent class analysis}.
\newblock
\APACaddressPublisher{Newby Park}{Sage}.
\PrintBackRefs{\CurrentBib}

\bibitem [\protect \citeauthoryear {%
Reid%
, Bycroft%
\BCBL {}\ \BBA {} Gleisner%
}{%
Reid%
\ \protect \BOthers {.}}{%
{\protect \APACyear {2016}}%
}]{%
ReidBycroftGleisner2016}
\APACinsertmetastar {%
ReidBycroftGleisner2016}%
\begin{APACrefauthors}%
Reid, G.%
, Bycroft, C.%
\BCBL {}\ \BBA {} Gleisner, F.%
\end{APACrefauthors}%
\unskip\
\newblock
\APACrefYear{2016}.
\newblock
\APACrefbtitle {Comparison of ethnicity information in administrative data and
  the census} {Comparison of ethnicity information in administrative data and
  the census}.
\newblock
\APACaddressPublisher{Christchurch}{Statistics New Zealand}.
\newblock
\begin{APACrefURL}
  \url{https://www.stats.govt.nz/assets/Research/Comparison-of-ethnicity-information-in-administrative-data-and-the-census/comparison-of-ethnicity-information-in-administrative-data-and-the-census.pdf}
  \end{APACrefURL}
\PrintBackRefs{\CurrentBib}

\bibitem [\protect \citeauthoryear {%
Simpson%
, Jivraj%
\BCBL {}\ \BBA {} Warren%
}{%
Simpson%
\ \protect \BOthers {.}}{%
{\protect \APACyear {2016}}%
}]{%
SimpsonLJivrajSWarrenJ2016}
\APACinsertmetastar {%
SimpsonLJivrajSWarrenJ2016}%
\begin{APACrefauthors}%
Simpson, L.%
, Jivraj, S.%
\BCBL {}\ \BBA {} Warren, J.%
\end{APACrefauthors}%
\unskip\
\newblock
\APACrefYearMonthDay{2016}{}{}.
\newblock
{\BBOQ}\APACrefatitle {The stability of ethnic identity in {E}ngland and
  {W}ales 2001–2011} {The stability of ethnic identity in {E}ngland and
  {W}ales 2001–2011}.{\BBCQ}
\newblock
\APACjournalVolNumPages{Journal of the Royal Statistical Society: Series
  A}{179}{}{1025--1049}.
\PrintBackRefs{\CurrentBib}

\bibitem [\protect \citeauthoryear {%
Stanghellini%
\ \BBA {} Van~der Heijden%
}{%
Stanghellini%
\ \BBA {} Van~der Heijden%
}{%
{\protect \APACyear {2004}}%
}]{%
StanghellinivanderHeijden2004}
\APACinsertmetastar {%
StanghellinivanderHeijden2004}%
\begin{APACrefauthors}%
Stanghellini, E.%
\BCBT {}\ \BBA {} Van~der Heijden, P\BPBI G\BPBI M.%
\end{APACrefauthors}%
\unskip\
\newblock
\APACrefYearMonthDay{2004}{}{}.
\newblock
{\BBOQ}\APACrefatitle {A Multiple-Record Systems Estimation Method that Takes
  Observed and Unobserved Heterogeneity into Account} {A multiple-record
  systems estimation method that takes observed and unobserved heterogeneity
  into account}.{\BBCQ}
\newblock
\APACjournalVolNumPages{Biometrics}{60}{2}{510-516}.
\newblock
\begin{APACrefDOI} \doi{10.1111/j.0006-341X.2004.00197.x} \end{APACrefDOI}
\PrintBackRefs{\CurrentBib}

\bibitem [\protect \citeauthoryear {%
{\noopsort{Statistics}}{Statistics New Zealand}%
}{%
{\noopsort{Statistics}}{Statistics New Zealand}%
}{%
{\protect \APACyear {2012}}%
}]{%
StatisticsNewZealand2012}
\APACinsertmetastar {%
StatisticsNewZealand2012}%
\begin{APACrefauthors}%
{\noopsort{Statistics}}{Statistics New Zealand}.%
\end{APACrefauthors}%
\unskip\
\newblock
\APACrefYear{2012}.
\newblock
\APACrefbtitle {Transforming the New Zealand Census of Population and
  Dwellings: Issues, options, and strategy} {Transforming the new zealand
  census of population and dwellings: Issues, options, and strategy}.
\newblock
\APACaddressPublisher{}{Christchurch, Statistics New Zealand}.
\newblock
\begin{APACrefURL} \url{https://www.stats.govt.nz} \end{APACrefURL}
\PrintBackRefs{\CurrentBib}

\bibitem [\protect \citeauthoryear {%
{\noopsort{Statistics}}{Statistics New Zealand}%
}{%
{\noopsort{Statistics}}{Statistics New Zealand}%
}{%
{\protect \APACyear {2014}}%
}]{%
StatisticsNewZealand2014}
\APACinsertmetastar {%
StatisticsNewZealand2014}%
\begin{APACrefauthors}%
{\noopsort{Statistics}}{Statistics New Zealand}.%
\end{APACrefauthors}%
\unskip\
\newblock
\APACrefYear{2014}.
\newblock
\APACrefbtitle {An overview of progress on the potential use of administrative
  data for census information in New Zealand: Census Transformation programme}
  {An overview of progress on the potential use of administrative data for
  census information in new zealand: Census transformation programme}.
\newblock
\APACaddressPublisher{}{Christchurch, Statistics New Zealand}.
\newblock
\begin{APACrefURL} \url{https://www.stats.govt.nz} \end{APACrefURL}
\PrintBackRefs{\CurrentBib}

\bibitem [\protect \citeauthoryear {%
{\noopsort{Statistics}}{Statistics New Zealand}%
}{%
{\noopsort{Statistics}}{Statistics New Zealand}%
}{%
{\protect \APACyear {2017}}%
{\protect \APACexlab {{\protect \BCnt {1}}}}}]{%
StatisticsNewZealand2017a}
\APACinsertmetastar {%
StatisticsNewZealand2017a}%
\begin{APACrefauthors}%
{\noopsort{Statistics}}{Statistics New Zealand}.%
\end{APACrefauthors}%
\unskip\
\newblock
\APACrefYear{2017{\protect \BCnt {1}}}.
\newblock
\APACrefbtitle {Experimental population estimates from linked administrative
  data: 2017 release} {Experimental population estimates from linked
  administrative data: 2017 release}.
\newblock
\APACaddressPublisher{}{Christchurch, Statistics New Zealand}.
\newblock
\begin{APACrefURL} \url{https://www.stats.govt.nz} \end{APACrefURL}
\PrintBackRefs{\CurrentBib}

\bibitem [\protect \citeauthoryear {%
{\noopsort{Statistics}}{Statistics New Zealand}%
}{%
{\noopsort{Statistics}}{Statistics New Zealand}%
}{%
{\protect \APACyear {2017}}%
{\protect \APACexlab {{\protect \BCnt {2}}}}}]{%
StatisticsNewZealand2017b}
\APACinsertmetastar {%
StatisticsNewZealand2017b}%
\begin{APACrefauthors}%
{\noopsort{Statistics}}{Statistics New Zealand}.%
\end{APACrefauthors}%
\unskip\
\newblock
\APACrefYear{2017{\protect \BCnt {2}}}.
\newblock
\APACrefbtitle {Integrated Data Infrastructure: Overarching privacy impact
  assessment} {Integrated data infrastructure: Overarching privacy impact
  assessment}.
\newblock
\APACaddressPublisher{}{Christchurch, Statistics New Zealand}.
\newblock
\begin{APACrefURL}
  \url{https://www.stats.govt.nz/assets/Uploads/Integrated-data-infrastructure/idi-overarching-pia.pdf}
  \end{APACrefURL}
\PrintBackRefs{\CurrentBib}

\bibitem [\protect \citeauthoryear {%
{\noopsort{Statistics}}{Statistics New Zealand}%
}{%
{\noopsort{Statistics}}{Statistics New Zealand}%
}{%
{\protect \APACyear {2018}}%
}]{%
StatisticsNewZealand2018}
\APACinsertmetastar {%
StatisticsNewZealand2018}%
\begin{APACrefauthors}%
{\noopsort{Statistics}}{Statistics New Zealand}.%
\end{APACrefauthors}%
\unskip\
\newblock
\APACrefYear{2018}.
\newblock
\APACrefbtitle {Experimental ethnic population estimates from linked
  administrative data} {Experimental ethnic population estimates from linked
  administrative data}.
\newblock
\APACaddressPublisher{}{Christchurch, Statistics New Zealand}.
\newblock
\begin{APACrefURL} \url{https://www.stats.govt.nz} \end{APACrefURL}
\PrintBackRefs{\CurrentBib}

\bibitem [\protect \citeauthoryear {%
Sutherland%
, Schwarz%
\BCBL {}\ \BBA {} Rivest%
}{%
Sutherland%
\ \protect \BOthers {.}}{%
{\protect \APACyear {2007}}%
}]{%
SutherlandJMSchwarzCJRivestLP2007}
\APACinsertmetastar {%
SutherlandJMSchwarzCJRivestLP2007}%
\begin{APACrefauthors}%
Sutherland, J.%
, Schwarz, C.%
\BCBL {}\ \BBA {} Rivest, L\BHBI P.%
\end{APACrefauthors}%
\unskip\
\newblock
\APACrefYearMonthDay{2007}{}{}.
\newblock
{\BBOQ}\APACrefatitle {Multilist population estimation with incomplete and
  partial stratification} {Multilist population estimation with incomplete and
  partial stratification}.{\BBCQ}
\newblock
\APACjournalVolNumPages{Biometrics}{63}{}{910-–916}.
\newblock
\begin{APACrefDOI} \doi{10.1111/j.1541-0420.2007.00767.x} \end{APACrefDOI}
\PrintBackRefs{\CurrentBib}

\bibitem [\protect \citeauthoryear {%
Van~der Heijden%
, Smith%
, Cruyff%
\BCBL {}\ \BBA {} Bakker%
}{%
Van~der Heijden%
\ \protect \BOthers {.}}{%
{\protect \APACyear {2018}}%
}]{%
VanderHeijdenPGMSmithPCruyffMBakkerB2018}
\APACinsertmetastar {%
VanderHeijdenPGMSmithPCruyffMBakkerB2018}%
\begin{APACrefauthors}%
Van~der Heijden, P\BPBI G\BPBI M.%
, Smith, P.%
, Cruyff, M.%
\BCBL {}\ \BBA {} Bakker, B.%
\end{APACrefauthors}%
\unskip\
\newblock
\APACrefYearMonthDay{2018}{}{}.
\newblock
{\BBOQ}\APACrefatitle {An overview of population size estimation where linking
  registers results in incomplete covariates, with an application to mode of
  transport of serious road casualties} {An overview of population size
  estimation where linking registers results in incomplete covariates, with an
  application to mode of transport of serious road casualties}.{\BBCQ}
\newblock
\APACjournalVolNumPages{Journal of Official Statistics}{34}{}{239--263}.
\PrintBackRefs{\CurrentBib}

\bibitem [\protect \citeauthoryear {%
Van~der Heijden%
\ \BBA {} Smith%
}{%
Van~der Heijden%
\ \BBA {} Smith%
}{%
{\protect \APACyear {2020}}%
}]{%
heijden2020estimating}
\APACinsertmetastar {%
heijden2020estimating}%
\begin{APACrefauthors}%
Van~der Heijden, P\BPBI G\BPBI M.%
\BCBT {}\ \BBA {} Smith, P\BPBI A.%
\end{APACrefauthors}%
\unskip\
\newblock
\APACrefYearMonthDay{2020}{}{}.
\newblock
\APACrefbtitle {On estimating the size of overcoverage with the latent class
  model. {A} critique of the paper "Population Size Estimation Using Multiple
  Incomplete Lists with Overcoverage" by di {C}ecco, di {Z}io, {F}ilipponi and
  {R}occhetti (2018, {JOS} 34 557-572).} {On estimating the size of
  overcoverage with the latent class model. {A} critique of the paper
  "population size estimation using multiple incomplete lists with
  overcoverage" by di {C}ecco, di {Z}io, {F}ilipponi and {R}occhetti (2018,
  {JOS} 34 557-572).}
\newblock
\APACrefnote{arXiv:2005.05452v1}
\PrintBackRefs{\CurrentBib}

\bibitem [\protect \citeauthoryear {%
Van~der Heijden%
, Whittaker%
, Cruyff%
, Bakker%
\BCBL {}\ \BBA {} Van~der Vliet%
}{%
Van~der Heijden%
\ \protect \BOthers {.}}{%
{\protect \APACyear {2012}}%
}]{%
VanderHeijdenPGMWhittakerCruyffBakkerVanderVliet2012}
\APACinsertmetastar {%
VanderHeijdenPGMWhittakerCruyffBakkerVanderVliet2012}%
\begin{APACrefauthors}%
Van~der Heijden, P\BPBI G\BPBI M.%
, Whittaker, J.%
, Cruyff, M.%
, Bakker, B.%
\BCBL {}\ \BBA {} Van~der Vliet, R.%
\end{APACrefauthors}%
\unskip\
\newblock
\APACrefYearMonthDay{2012}{}{}.
\newblock
{\BBOQ}\APACrefatitle {People born in the {M}iddle {E}ast but residing in the
  {N}etherlands: Invariant population size estimates and the role of active and
  passive covariates} {People born in the {M}iddle {E}ast but residing in the
  {N}etherlands: Invariant population size estimates and the role of active and
  passive covariates}.{\BBCQ}
\newblock
\APACjournalVolNumPages{The Annals of Applied Statistics}{6}{}{831--852}.
\newblock
\begin{APACrefDOI} \doi{10.1214/12-AOAS536} \end{APACrefDOI}
\PrintBackRefs{\CurrentBib}

\bibitem [\protect \citeauthoryear {%
Vermunt%
\ \BBA {} Magidson%
}{%
Vermunt%
\ \BBA {} Magidson%
}{%
{\protect \APACyear {2004}}%
}]{%
VermuntMagidson2004}
\APACinsertmetastar {%
VermuntMagidson2004}%
\begin{APACrefauthors}%
Vermunt, J\BPBI K.%
\BCBT {}\ \BBA {} Magidson, J.%
\end{APACrefauthors}%
\unskip\
\newblock
\APACrefYearMonthDay{2004}{}{}.
\newblock
{\BBOQ}\APACrefatitle {Latent class analysis} {Latent class analysis}.{\BBCQ}
\newblock
\BIn{} M\BPBI S.~Lewis-Beck, A.~A.~Bryman\BCBL {}\ \BBA {} T.~Liao\ (\BEDS),
  \APACrefbtitle {The Sage Encyclopedia of Social Sciences Research Methods}
  {The sage encyclopedia of social sciences research methods}\ (\BPG~549-553).
\newblock
\APACaddressPublisher{Thousand Oaks, CA}{Sage Publications}.
\PrintBackRefs{\CurrentBib}

\bibitem [\protect \citeauthoryear {%
Waldon%
}{%
Waldon%
}{%
{\protect \APACyear {2019}}%
}]{%
WaldonJ2019}
\APACinsertmetastar {%
WaldonJ2019}%
\begin{APACrefauthors}%
Waldon, J.%
\end{APACrefauthors}%
\unskip\
\newblock
\APACrefYearMonthDay{2019}{}{}.
\newblock
{\BBOQ}\APACrefatitle {Identification of indigenous people in A otearoa-{N}ew
  {Z}ealand – {N}gā mata o taku {W}henua} {Identification of indigenous
  people in a otearoa-{N}ew {Z}ealand – {N}gā mata o taku {W}henua}.{\BBCQ}
\newblock
\APACjournalVolNumPages{Statistical Journal of the IAOS}{35}{}{107--118}.
\PrintBackRefs{\CurrentBib}

\bibitem [\protect \citeauthoryear {%
Whittaker%
}{%
Whittaker%
}{%
{\protect \APACyear {1990}}%
}]{%
WhittakerJ1990}
\APACinsertmetastar {%
WhittakerJ1990}%
\begin{APACrefauthors}%
Whittaker, J.%
\end{APACrefauthors}%
\unskip\
\newblock
\APACrefYear{1990}.
\newblock
\APACrefbtitle {Graphical models in applied multivariate statistics} {Graphical
  models in applied multivariate statistics}.
\newblock
\APACaddressPublisher{Chichester}{Wiley}.
\PrintBackRefs{\CurrentBib}

\bibitem [\protect \citeauthoryear {%
Wolter%
}{%
Wolter%
}{%
{\protect \APACyear {1986}}%
}]{%
WolterKM1986}
\APACinsertmetastar {%
WolterKM1986}%
\begin{APACrefauthors}%
Wolter, K\BPBI M.%
\end{APACrefauthors}%
\unskip\
\newblock
\APACrefYearMonthDay{1986}{}{}.
\newblock
{\BBOQ}\APACrefatitle {Some coverage error models for census data} {Some
  coverage error models for census data}.{\BBCQ}
\newblock
\APACjournalVolNumPages{Journal of the American Statistical
  Association}{81}{394}{337--346}.
\newblock
\begin{APACrefDOI} \doi{10.1080/01621459.1986.10478277} \end{APACrefDOI}
\PrintBackRefs{\CurrentBib}

\bibitem [\protect \citeauthoryear {%
Zhang%
\ \BBA {} Chambers%
}{%
Zhang%
\ \BBA {} Chambers%
}{%
{\protect \APACyear {2019}}%
}]{%
ZhangChambers2019}
\APACinsertmetastar {%
ZhangChambers2019}%
\begin{APACrefauthors}%
Zhang, L\BHBI C.%
\BCBT {}\ \BBA {} Chambers, R\BPBI L.%
\end{APACrefauthors}%
\unskip\
\newblock
\APACrefYear{2019}.
\newblock
\APACrefbtitle {Analysis of integrated data} {Analysis of integrated data}.
\newblock
\APACaddressPublisher{Boca Raton}{CRC Press}.
\PrintBackRefs{\CurrentBib}

\bibitem [\protect \citeauthoryear {%
Zhang%
\ \BBA {} Dunne%
}{%
Zhang%
\ \BBA {} Dunne%
}{%
{\protect \APACyear {2018}}%
}]{%
ZhangDunne2018}
\APACinsertmetastar {%
ZhangDunne2018}%
\begin{APACrefauthors}%
Zhang, L\BHBI C.%
\BCBT {}\ \BBA {} Dunne, J.%
\end{APACrefauthors}%
\unskip\
\newblock
\APACrefYearMonthDay{2018}{}{}.
\newblock
{\BBOQ}\APACrefatitle {Trimmed dual system estimation} {Trimmed dual system
  estimation}.{\BBCQ}
\newblock
\BIn{} D.~B\"{o}hning, P.~Van~der Heijden\BCBL {}\ \BBA {} J.~Bunge\ (\BEDS),
  \APACrefbtitle {Capture-recapture methods for the social and medical
  sciences} {Capture-recapture methods for the social and medical sciences}\
  (\BPGS\ 229--235).
\newblock
\APACaddressPublisher{Boca Raton}{CRC Press}.
\PrintBackRefs{\CurrentBib}

\bibitem [\protect \citeauthoryear {%
Zwane%
\ \BBA {} {van der Heijden}%
}{%
Zwane%
\ \BBA {} {van der Heijden}%
}{%
{\protect \APACyear {2007}}%
}]{%
ZwaneEvanderHeijdenPGM2007}
\APACinsertmetastar {%
ZwaneEvanderHeijdenPGM2007}%
\begin{APACrefauthors}%
Zwane, E.%
\BCBT {}\ \BBA {} {van der Heijden}, P\BPBI G\BPBI M.%
\end{APACrefauthors}%
\unskip\
\newblock
\APACrefYearMonthDay{2007}{}{}.
\newblock
{\BBOQ}\APACrefatitle {Analysing capture-recapture data when some variables of
  heterogeneous catchability are not collected or asked in all registrations}
  {Analysing capture-recapture data when some variables of heterogeneous
  catchability are not collected or asked in all registrations}.{\BBCQ}
\newblock
\APACjournalVolNumPages{Statistics in Medicine}{26}{}{1069-1089}.
\newblock
\begin{APACrefDOI} \doi{10.1002/sim.2577} \end{APACrefDOI}
\PrintBackRefs{\CurrentBib}

\bibitem [\protect \citeauthoryear {%
Zwane%
, {Van der Pal-de Bruin}%
\BCBL {}\ \BBA {} {Van der Heijden}%
}{%
Zwane%
\ \protect \BOthers {.}}{%
{\protect \APACyear {2004}}%
}]{%
ZwaneEvanderPaldeBruinKvanderHeijdenPGM2004}
\APACinsertmetastar {%
ZwaneEvanderPaldeBruinKvanderHeijdenPGM2004}%
\begin{APACrefauthors}%
Zwane, E.%
, {Van der Pal-de Bruin}, K.%
\BCBL {}\ \BBA {} {Van der Heijden}, P\BPBI G\BPBI M.%
\end{APACrefauthors}%
\unskip\
\newblock
\APACrefYearMonthDay{2004}{}{}.
\newblock
{\BBOQ}\APACrefatitle {The multiple-record systems estimator when registrations
  refer to different but overlapping populations} {The multiple-record systems
  estimator when registrations refer to different but overlapping
  populations}.{\BBCQ}
\newblock
\APACjournalVolNumPages{Statistics in Medicine}{23}{}{2267-2281}.
\newblock
\begin{APACrefDOI} \doi{10.1002/sim.1818} \end{APACrefDOI}
\PrintBackRefs{\CurrentBib}

\end{thebibliography}

\end{document}


\noindent\textbf{\large{Supplementary Material A for “Multiple system estimation using covariates having missing values and measurement error: estimating the size of the M\=aori population in New Zealand”}} by Peter G.M. van der Heijden, Maarten Cruyff, Paul A. Smith, Christine Bycroft, Patrick Graham and Nathaniel Matheson-Dunning

\vspace{5mm}





\noindent\textbf{S1: Raw data tables}
\vspace{5mm}

\noindent In principle all the tables can be derived by aggregation from the four-way ABCD table (Table S3 below), but all of the individual tables which are the basis of the analyses in sections 4.1-4.4 are provided here for convenience. The original data were randomly rounded to base 3 to protect confidentiality. Profiles with an observed count of zero are not displayed. The data are available at \url{https://github.com/MaartenCruyff/cenzus}.
\vspace{5mm}

\noindent Table S1: Table of counts from Census (A) linked to MOH (C). Ethnicity in the registers is coded with lowercase letters, taking values 0 (non-Māori), 1 (Māori), and ‒ (missing).

\begin{center}
\section*{AC}\label{ac}

\begin{tabular}{l l l l r}
\hline
A & C & a & c & Freq\\
\hline
1 & 1 & 0 & 0 & 3004335\\
1 & 1 & 0 & 1 & 31995\\
1 & 0 & 0 & ‒ & 38634\\
1 & 1 & 0 & ‒ & 150840\\
1 & 1 & 1 & 0 & 108189\\
1 & 1 & 1 & 1 & 435465\\
1 & 0 & 1 & ‒ & 4368\\
1 & 1 & 1 & ‒ & 12405\\
0 & 1 & ‒ & 0 & 398838\\
1 & 1 & ‒ & 0 & 16512\\
0 & 1 & ‒ & 1 & 146976\\
1 & 1 & ‒ & 1 & 2769\\
1 & 0 & ‒ & ‒ & 438\\
0 & 1 & ‒ & ‒ & 24636\\
1 & 1 & ‒ & ‒ & 900\\
\hline
\end{tabular}
\end{center}

Source: Stats NZ
\newpage

\noindent Table S2: Table of counts from linkage of Census (A), DIA (B) and MOH (C) registers. Ethnicity in the registers is coded with lowercase letters, taking values 0 (non-Māori), 1 (Māori), and ‒ (missing).

\begin{center}

\section*{ABC}\label{abc}

\begin{longtable}{l l l l l l r}
\hline
A & B & C & a & b & c & Freq\\
\hline
1 & 1 & 1 & 0 & 0 & 0 & 497148\\
1 & 1 & 1 & 0 & 0 & 1 & 4530\\
1 & 1 & 0 & 0 & 0 & ‒ & 927\\
1 & 1 & 1 & 0 & 0 & ‒ & 3990\\
1 & 1 & 1 & 0 & 1 & 0 & 8493\\
1 & 1 & 1 & 0 & 1 & 1 & 6225\\
1 & 1 & 0 & 0 & 1 & ‒ & 33\\
1 & 1 & 1 & 0 & 1 & ‒ & 69\\
1 & 0 & 1 & 0 & ‒ & 0 & 2494896\\
1 & 1 & 1 & 0 & ‒ & 0 & 3798\\
1 & 0 & 1 & 0 & ‒ & 1 & 21150\\
1 & 1 & 1 & 0 & ‒ & 1 & 90\\
1 & 0 & 0 & 0 & ‒ & ‒ & 37650\\
1 & 1 & 0 & 0 & ‒ & ‒ & 24\\
1 & 0 & 1 & 0 & ‒ & ‒ & 146721\\
1 & 1 & 1 & 0 & ‒ & ‒ & 60\\
1 & 1 & 1 & 1 & 0 & 0 & 9873\\
1 & 1 & 1 & 1 & 0 & 1 & 4662\\
1 & 1 & 0 & 1 & 0 & ‒ & 24\\
1 & 1 & 1 & 1 & 0 & ‒ & 84\\
1 & 1 & 1 & 1 & 1 & 0 & 31266\\
1 & 1 & 1 & 1 & 1 & 1 & 140250\\
1 & 1 & 0 & 1 & 1 & ‒ & 441\\
1 & 1 & 1 & 1 & 1 & ‒ & 621\\
1 & 0 & 1 & 1 & ‒ & 0 & 66762\\
1 & 1 & 1 & 1 & ‒ & 0 & 288\\
1 & 0 & 1 & 1 & ‒ & 1 & 289770\\
1 & 1 & 1 & 1 & ‒ & 1 & 783\\
1 & 0 & 0 & 1 & ‒ & ‒ & 3903\\
1 & 0 & 1 & 1 & ‒ & ‒ & 11691\\
1 & 1 & 1 & 1 & ‒ & ‒ & 9\\
0 & 1 & 1 & ‒ & 0 & 0 & 47754\\
1 & 1 & 1 & ‒ & 0 & 0 & 1923\\
0 & 1 & 1 & ‒ & 0 & 1 & 2160\\
1 & 1 & 1 & ‒ & 0 & 1 & 45\\
0 & 1 & 0 & ‒ & 0 & ‒ & 684\\
0 & 1 & 1 & ‒ & 0 & ‒ & 255\\
1 & 1 & 1 & ‒ & 0 & ‒ & 18\\
0 & 1 & 1 & ‒ & 1 & 0 & 5751\\
1 & 1 & 1 & ‒ & 1 & 0 & 156\\
0 & 1 & 1 & ‒ & 1 & 1 & 41952\\
1 & 1 & 1 & ‒ & 1 & 1 & 936\\
0 & 1 & 0 & ‒ & 1 & ‒ & 375\\
0 & 1 & 1 & ‒ & 1 & ‒ & 105\\
1 & 0 & 1 & ‒ & ‒ & 0 & 14349\\
0 & 1 & 1 & ‒ & ‒ & 0 & 603\\
1 & 1 & 1 & ‒ & ‒ & 0 & 84\\
1 & 0 & 1 & ‒ & ‒ & 1 & 1779\\
0 & 1 & 1 & ‒ & ‒ & 1 & 273\\
1 & 1 & 1 & ‒ & ‒ & 1 & 9\\
0 & 0 & 0 & ‒ & ‒ & ‒ & 23613\\
1 & 0 & 0 & ‒ & ‒ & ‒ & 438\\
0 & 1 & 0 & ‒ & ‒ & ‒ & 18\\
1 & 0 & 1 & ‒ & ‒ & ‒ & 882\\
0 & 1 & 1 & ‒ & ‒ & ‒ & 6\\
\hline
\end{longtable}
\end{center}

Source: Stats NZ
\newpage

\noindent Table S3: Table of counts from linkage of Census (A), DIA (B), MOH (C) and MOE (D) registers. Ethnicity in the registers is coded with lowercase letters, taking values 0 (non-Māori), 1 (Māori), and ‒ (missing).
\begin{center}
\section*{ABCD}\label{abcd}
\begin{longtable}{r r r r r r r r r}
\hline
A & B & C & D & a & b & c & d & Freq\\
\hline
1 & 1 & 1 & 1 & 0 & 0 & 0 & 0 & 44448\\
1 & 1 & 1 & 1 & 0 & 0 & 0 & 1 & 1053\\
1 & 1 & 1 & 0 & 0 & 0 & 0 & ‒ & 450582\\
1 & 1 & 1 & 1 & 0 & 0 & 0 & ‒ & 1065\\
1 & 1 & 1 & 1 & 0 & 0 & 1 & 0 & 246\\
1 & 1 & 1 & 1 & 0 & 0 & 1 & 1 & 201\\
1 & 1 & 1 & 0 & 0 & 0 & 1 & ‒ & 4071\\
1 & 1 & 1 & 1 & 0 & 0 & 1 & ‒ & 12\\
1 & 1 & 0 & 1 & 0 & 0 & ‒ & 0 & 117\\
1 & 1 & 1 & 1 & 0 & 0 & ‒ & 0 & 789\\
1 & 1 & 1 & 1 & 0 & 0 & ‒ & 1 & 9\\
1 & 1 & 0 & 0 & 0 & 0 & ‒ & ‒ & 810\\
1 & 1 & 1 & 0 & 0 & 0 & ‒ & ‒ & 3174\\
1 & 1 & 1 & 1 & 0 & 0 & ‒ & ‒ & 18\\
1 & 1 & 1 & 1 & 0 & 1 & 0 & 0 & 531\\
1 & 1 & 1 & 1 & 0 & 1 & 0 & 1 & 315\\
1 & 1 & 1 & 0 & 0 & 1 & 0 & ‒ & 7626\\
1 & 1 & 1 & 1 & 0 & 1 & 0 & ‒ & 21\\
1 & 1 & 1 & 1 & 0 & 1 & 1 & 0 & 195\\
1 & 1 & 1 & 1 & 0 & 1 & 1 & 1 & 507\\
1 & 1 & 1 & 0 & 0 & 1 & 1 & ‒ & 5511\\
1 & 1 & 1 & 1 & 0 & 1 & 1 & ‒ & 12\\
1 & 1 & 1 & 1 & 0 & 1 & ‒ & 1 & 9\\
1 & 1 & 0 & 0 & 0 & 1 & ‒ & ‒ & 33\\
1 & 1 & 1 & 0 & 0 & 1 & ‒ & ‒ & 60\\
1 & 0 & 1 & 1 & 0 & ‒ & 0 & 0 & 1364184\\
1 & 1 & 1 & 1 & 0 & ‒ & 0 & 0 & 492\\
1 & 0 & 1 & 1 & 0 & ‒ & 0 & 1 & 22554\\
1 & 1 & 1 & 1 & 0 & ‒ & 0 & 1 & 21\\
1 & 0 & 1 & 0 & 0 & ‒ & 0 & ‒ & 1095207\\
1 & 1 & 1 & 0 & 0 & ‒ & 0 & ‒ & 3267\\
1 & 0 & 1 & 1 & 0 & ‒ & 0 & ‒ & 12951\\
1 & 1 & 1 & 1 & 0 & ‒ & 0 & ‒ & 18\\
1 & 0 & 1 & 1 & 0 & ‒ & 1 & 0 & 6399\\
1 & 1 & 1 & 1 & 0 & ‒ & 1 & 0 & 9\\
1 & 0 & 1 & 1 & 0 & ‒ & 1 & 1 & 7452\\
1 & 1 & 1 & 1 & 0 & ‒ & 1 & 1 & 9\\
1 & 0 & 1 & 0 & 0 & ‒ & 1 & ‒ & 7197\\
1 & 1 & 1 & 0 & 0 & ‒ & 1 & ‒ & 72\\
1 & 0 & 1 & 1 & 0 & ‒ & 1 & ‒ & 102\\
1 & 0 & 0 & 1 & 0 & ‒ & ‒ & 0 & 15882\\
1 & 0 & 1 & 1 & 0 & ‒ & ‒ & 0 & 80025\\
1 & 1 & 1 & 1 & 0 & ‒ & ‒ & 0 & 12\\
1 & 0 & 0 & 1 & 0 & ‒ & ‒ & 1 & 207\\
1 & 0 & 1 & 1 & 0 & ‒ & ‒ & 1 & 1320\\
1 & 0 & 0 & 0 & 0 & ‒ & ‒ & ‒ & 21135\\
1 & 1 & 0 & 0 & 0 & ‒ & ‒ & ‒ & 24\\
1 & 0 & 1 & 0 & 0 & ‒ & ‒ & ‒ & 64635\\
1 & 1 & 1 & 0 & 0 & ‒ & ‒ & ‒ & 48\\
1 & 0 & 0 & 1 & 0 & ‒ & ‒ & ‒ & 426\\
1 & 0 & 1 & 1 & 0 & ‒ & ‒ & ‒ & 741\\
1 & 1 & 1 & 1 & 1 & 0 & 0 & 0 & 621\\
1 & 1 & 1 & 1 & 1 & 0 & 0 & 1 & 771\\
1 & 1 & 1 & 0 & 1 & 0 & 0 & ‒ & 8448\\
1 & 1 & 1 & 1 & 1 & 0 & 0 & ‒ & 33\\
1 & 1 & 1 & 1 & 1 & 0 & 1 & 0 & 147\\
1 & 1 & 1 & 1 & 1 & 0 & 1 & 1 & 579\\
1 & 1 & 1 & 0 & 1 & 0 & 1 & ‒ & 3918\\
1 & 1 & 1 & 1 & 1 & 0 & 1 & ‒ & 18\\
1 & 1 & 1 & 1 & 1 & 0 & ‒ & 0 & 6\\
1 & 1 & 1 & 1 & 1 & 0 & ‒ & 1 & 15\\
1 & 1 & 0 & 0 & 1 & 0 & ‒ & ‒ & 24\\
1 & 1 & 1 & 0 & 1 & 0 & ‒ & ‒ & 63\\
1 & 1 & 1 & 1 & 1 & 1 & 0 & 0 & 564\\
1 & 1 & 1 & 1 & 1 & 1 & 0 & 1 & 2232\\
1 & 1 & 1 & 0 & 1 & 1 & 0 & ‒ & 28398\\
1 & 1 & 1 & 1 & 1 & 1 & 0 & ‒ & 72\\
1 & 1 & 1 & 1 & 1 & 1 & 1 & 0 & 1296\\
1 & 1 & 1 & 1 & 1 & 1 & 1 & 1 & 14196\\
1 & 1 & 1 & 0 & 1 & 1 & 1 & ‒ & 124515\\
1 & 1 & 1 & 1 & 1 & 1 & 1 & ‒ & 243\\
1 & 1 & 0 & 1 & 1 & 1 & ‒ & 0 & 15\\
1 & 1 & 1 & 1 & 1 & 1 & ‒ & 0 & 9\\
1 & 1 & 0 & 1 & 1 & 1 & ‒ & 1 & 81\\
1 & 1 & 1 & 1 & 1 & 1 & ‒ & 1 & 126\\
1 & 1 & 0 & 0 & 1 & 1 & ‒ & ‒ & 345\\
1 & 1 & 1 & 0 & 1 & 1 & ‒ & ‒ & 486\\
1 & 0 & 1 & 1 & 1 & ‒ & 0 & 0 & 11382\\
1 & 1 & 1 & 1 & 1 & ‒ & 0 & 0 & 15\\
1 & 0 & 1 & 1 & 1 & ‒ & 0 & 1 & 35292\\
1 & 1 & 1 & 1 & 1 & ‒ & 0 & 1 & 36\\
1 & 0 & 1 & 0 & 1 & ‒ & 0 & ‒ & 19737\\
1 & 1 & 1 & 0 & 1 & ‒ & 0 & ‒ & 237\\
1 & 0 & 1 & 1 & 1 & ‒ & 0 & ‒ & 351\\
1 & 0 & 1 & 1 & 1 & ‒ & 1 & 0 & 8298\\
1 & 1 & 1 & 1 & 1 & ‒ & 1 & 0 & 12\\
1 & 0 & 1 & 1 & 1 & ‒ & 1 & 1 & 210696\\
1 & 1 & 1 & 1 & 1 & ‒ & 1 & 1 & 171\\
1 & 0 & 1 & 0 & 1 & ‒ & 1 & ‒ & 69768\\
1 & 1 & 1 & 0 & 1 & ‒ & 1 & ‒ & 600\\
1 & 0 & 1 & 1 & 1 & ‒ & 1 & ‒ & 1008\\
1 & 0 & 0 & 1 & 1 & ‒ & ‒ & 0 & 123\\
1 & 0 & 1 & 1 & 1 & ‒ & ‒ & 0 & 936\\
1 & 0 & 0 & 1 & 1 & ‒ & ‒ & 1 & 1920\\
1 & 0 & 1 & 1 & 1 & ‒ & ‒ & 1 & 7119\\
1 & 0 & 0 & 0 & 1 & ‒ & ‒ & ‒ & 1848\\
1 & 0 & 1 & 0 & 1 & ‒ & ‒ & ‒ & 3594\\
1 & 1 & 1 & 0 & 1 & ‒ & ‒ & ‒ & 9\\
1 & 0 & 0 & 1 & 1 & ‒ & ‒ & ‒ & 12\\
1 & 0 & 1 & 1 & 1 & ‒ & ‒ & ‒ & 42\\
0 & 1 & 1 & 1 & ‒ & 0 & 0 & 0 & 3687\\
1 & 1 & 1 & 1 & ‒ & 0 & 0 & 0 & 159\\
0 & 1 & 1 & 1 & ‒ & 0 & 0 & 1 & 243\\
1 & 1 & 1 & 1 & ‒ & 0 & 0 & 1 & 6\\
0 & 1 & 1 & 0 & ‒ & 0 & 0 & ‒ & 43722\\
1 & 1 & 1 & 0 & ‒ & 0 & 0 & ‒ & 1758\\
0 & 1 & 1 & 1 & ‒ & 0 & 0 & ‒ & 102\\
0 & 1 & 1 & 1 & ‒ & 0 & 1 & 0 & 78\\
0 & 1 & 1 & 1 & ‒ & 0 & 1 & 1 & 141\\
1 & 1 & 1 & 1 & ‒ & 0 & 1 & 1 & 6\\
0 & 1 & 1 & 0 & ‒ & 0 & 1 & ‒ & 1941\\
1 & 1 & 1 & 0 & ‒ & 0 & 1 & ‒ & 39\\
0 & 1 & 0 & 1 & ‒ & 0 & ‒ & 0 & 48\\
0 & 1 & 1 & 1 & ‒ & 0 & ‒ & 0 & 45\\
0 & 1 & 0 & 1 & ‒ & 0 & ‒ & 1 & 6\\
0 & 1 & 0 & 0 & ‒ & 0 & ‒ & ‒ & 630\\
0 & 1 & 1 & 0 & ‒ & 0 & ‒ & ‒ & 210\\
1 & 1 & 1 & 0 & ‒ & 0 & ‒ & ‒ & 18\\
0 & 1 & 1 & 1 & ‒ & 1 & 0 & 0 & 162\\
0 & 1 & 1 & 1 & ‒ & 1 & 0 & 1 & 387\\
1 & 1 & 1 & 1 & ‒ & 1 & 0 & 1 & 6\\
0 & 1 & 1 & 0 & ‒ & 1 & 0 & ‒ & 5187\\
1 & 1 & 1 & 0 & ‒ & 1 & 0 & ‒ & 150\\
0 & 1 & 1 & 1 & ‒ & 1 & 0 & ‒ & 15\\
0 & 1 & 1 & 1 & ‒ & 1 & 1 & 0 & 375\\
1 & 1 & 1 & 1 & ‒ & 1 & 1 & 0 & 12\\
0 & 1 & 1 & 1 & ‒ & 1 & 1 & 1 & 4071\\
1 & 1 & 1 & 1 & ‒ & 1 & 1 & 1 & 87\\
0 & 1 & 1 & 0 & ‒ & 1 & 1 & ‒ & 37458\\
1 & 1 & 1 & 0 & ‒ & 1 & 1 & ‒ & 837\\
0 & 1 & 1 & 1 & ‒ & 1 & 1 & ‒ & 48\\
0 & 1 & 0 & 1 & ‒ & 1 & ‒ & 1 & 60\\
0 & 1 & 1 & 1 & ‒ & 1 & ‒ & 1 & 24\\
0 & 1 & 0 & 0 & ‒ & 1 & ‒ & ‒ & 315\\
0 & 1 & 1 & 0 & ‒ & 1 & ‒ & ‒ & 81\\
0 & 0 & 1 & 1 & ‒ & ‒ & 0 & 0 & 181839\\
1 & 0 & 1 & 1 & ‒ & ‒ & 0 & 0 & 6093\\
0 & 1 & 1 & 1 & ‒ & ‒ & 0 & 0 & 54\\
0 & 0 & 1 & 1 & ‒ & ‒ & 0 & 1 & 12669\\
1 & 0 & 1 & 1 & ‒ & ‒ & 0 & 1 & 285\\
0 & 1 & 1 & 1 & ‒ & ‒ & 0 & 1 & 9\\
0 & 0 & 1 & 0 & ‒ & ‒ & 0 & ‒ & 148110\\
1 & 0 & 1 & 0 & ‒ & ‒ & 0 & ‒ & 7890\\
0 & 1 & 1 & 0 & ‒ & ‒ & 0 & ‒ & 540\\
1 & 1 & 1 & 0 & ‒ & ‒ & 0 & ‒ & 84\\
0 & 0 & 1 & 1 & ‒ & ‒ & 0 & ‒ & 2112\\
1 & 0 & 1 & 1 & ‒ & ‒ & 0 & ‒ & 81\\
0 & 0 & 1 & 1 & ‒ & ‒ & 1 & 0 & 3987\\
1 & 0 & 1 & 1 & ‒ & ‒ & 1 & 0 & 93\\
0 & 0 & 1 & 1 & ‒ & ‒ & 1 & 1 & 72747\\
1 & 0 & 1 & 1 & ‒ & ‒ & 1 & 1 & 1110\\
0 & 1 & 1 & 1 & ‒ & ‒ & 1 & 1 & 42\\
0 & 0 & 1 & 0 & ‒ & ‒ & 1 & ‒ & 25521\\
1 & 0 & 1 & 0 & ‒ & ‒ & 1 & ‒ & 570\\
0 & 1 & 1 & 0 & ‒ & ‒ & 1 & ‒ & 231\\
1 & 1 & 1 & 0 & ‒ & ‒ & 1 & ‒ & 9\\
0 & 0 & 1 & 1 & ‒ & ‒ & 1 & ‒ & 336\\
1 & 0 & 1 & 1 & ‒ & ‒ & 1 & ‒ & 6\\
0 & 0 & 0 & 1 & ‒ & ‒ & ‒ & 0 & 19368\\
1 & 0 & 0 & 1 & ‒ & ‒ & ‒ & 0 & 117\\
0 & 0 & 1 & 1 & ‒ & ‒ & ‒ & 0 & 10191\\
1 & 0 & 1 & 1 & ‒ & ‒ & ‒ & 0 & 402\\
0 & 0 & 0 & 1 & ‒ & ‒ & ‒ & 1 & 3786\\
1 & 0 & 0 & 1 & ‒ & ‒ & ‒ & 1 & 24\\
0 & 0 & 1 & 1 & ‒ & ‒ & ‒ & 1 & 2415\\
1 & 0 & 1 & 1 & ‒ & ‒ & ‒ & 1 & 48\\
1 & 0 & 0 & 0 & ‒ & ‒ & ‒ & ‒ & 291\\
0 & 1 & 0 & 0 & ‒ & ‒ & ‒ & ‒ & 18\\
0 & 0 & 1 & 0 & ‒ & ‒ & ‒ & ‒ & 11556\\
1 & 0 & 1 & 0 & ‒ & ‒ & ‒ & ‒ & 426\\
0 & 1 & 1 & 0 & ‒ & ‒ & ‒ & ‒ & 6\\
0 & 0 & 0 & 1 & ‒ & ‒ & ‒ & ‒ & 459\\
1 & 0 & 0 & 1 & ‒ & ‒ & ‒ & ‒ & 6\\
0 & 0 & 1 & 1 & ‒ & ‒ & ‒ & ‒ & 108\\
1 & 0 & 1 & 1 & ‒ & ‒ & ‒ & ‒ & 6\\
\hline
\end{longtable}    
\end{center}

Source: Stats NZ
\newpage

\noindent Table S4: Table of counts from linkage of DIA (B), MOH (C) and MOE (D) registers. Ethnicity in the registers is coded with lowercase letters, taking values 0 (non-Māori), 1 (Māori), and ‒ (missing).

\begin{center}
\section*{BCD}\label{bcd}

\begin{longtable}{l l l l l l r}
\hline
B & C & D & b & c & d & Freq\\
\hline
1 & 1 & 1 & 0 & 0 & 0 & 48915\\
1 & 1 & 1 & 0 & 0 & 1 & 2073\\
1 & 1 & 0 & 0 & 0 & ‒ & 504510\\
1 & 1 & 1 & 0 & 0 & ‒ & 1200\\
1 & 1 & 1 & 0 & 1 & 0 & 471\\
1 & 1 & 1 & 0 & 1 & 1 & 927\\
1 & 1 & 0 & 0 & 1 & ‒ & 9969\\
1 & 1 & 1 & 0 & 1 & ‒ & 30\\
1 & 0 & 1 & 0 & ‒ & 0 & 165\\
1 & 1 & 1 & 0 & ‒ & 0 & 840\\
1 & 0 & 1 & 0 & ‒ & 1 & 6\\
1 & 1 & 1 & 0 & ‒ & 1 & 24\\
1 & 0 & 0 & 0 & ‒ & ‒ & 1464\\
1 & 1 & 0 & 0 & ‒ & ‒ & 3465\\
1 & 1 & 1 & 0 & ‒ & ‒ & 18\\
1 & 1 & 1 & 1 & 0 & 0 & 1257\\
1 & 1 & 1 & 1 & 0 & 1 & 2940\\
1 & 1 & 0 & 1 & 0 & ‒ & 41361\\
1 & 1 & 1 & 1 & 0 & ‒ & 108\\
1 & 1 & 1 & 1 & 1 & 0 & 1878\\
1 & 1 & 1 & 1 & 1 & 1 & 18861\\
1 & 1 & 0 & 1 & 1 & ‒ & 168321\\
1 & 1 & 1 & 1 & 1 & ‒ & 303\\
1 & 0 & 1 & 1 & ‒ & 0 & 15\\
1 & 1 & 1 & 1 & ‒ & 0 & 9\\
1 & 0 & 1 & 1 & ‒ & 1 & 141\\
1 & 1 & 1 & 1 & ‒ & 1 & 159\\
1 & 0 & 0 & 1 & ‒ & ‒ & 693\\
1 & 1 & 0 & 1 & ‒ & ‒ & 627\\
0 & 1 & 1 & ‒ & 0 & 0 & 1563498\\
1 & 1 & 1 & ‒ & 0 & 0 & 561\\
0 & 1 & 1 & ‒ & 0 & 1 & 70800\\
1 & 1 & 1 & ‒ & 0 & 1 & 66\\
0 & 1 & 0 & ‒ & 0 & ‒ & 1270944\\
1 & 1 & 0 & ‒ & 0 & ‒ & 4128\\
0 & 1 & 1 & ‒ & 0 & ‒ & 15495\\
1 & 1 & 1 & ‒ & 0 & ‒ & 18\\
0 & 1 & 1 & ‒ & 1 & 0 & 18777\\
1 & 1 & 1 & ‒ & 1 & 0 & 21\\
0 & 1 & 1 & ‒ & 1 & 1 & 292005\\
1 & 1 & 1 & ‒ & 1 & 1 & 222\\
0 & 1 & 0 & ‒ & 1 & ‒ & 103056\\
1 & 1 & 0 & ‒ & 1 & ‒ & 912\\
0 & 1 & 1 & ‒ & 1 & ‒ & 1452\\
0 & 0 & 1 & ‒ & ‒ & 0 & 35490\\
0 & 1 & 1 & ‒ & ‒ & 0 & 91554\\
1 & 1 & 1 & ‒ & ‒ & 0 & 12\\
0 & 0 & 1 & ‒ & ‒ & 1 & 5937\\
0 & 1 & 1 & ‒ & ‒ & 1 & 10902\\
1 & 0 & 0 & ‒ & ‒ & ‒ & 42\\
0 & 1 & 0 & ‒ & ‒ & ‒ & 80211\\
1 & 1 & 0 & ‒ & ‒ & ‒ & 63\\
0 & 0 & 1 & ‒ & ‒ & ‒ & 903\\
0 & 1 & 1 & ‒ & ‒ & ‒ & 897\\
\hline
\end{longtable}
\end{center}

Source: Stats NZ

\newpage

\noindent\textbf{S2: Parameter estimates}
\vspace{5mm}

\noindent The parameter estimates from the two models for the four-register case examined in section 4.4 of the main paper are given below. Further output is available in ‘Supplementary Material B.html ’.
\vspace{5mm}

\noindent Table S5: The parameter estimates from the four-register analysis using the maximal model [ABCd][ABDc][ACDb] [BCDa][ABcd][ACbd][ADbc][BCad][BDac][CDab][Abcd][Bacd]
\newline[Cabd][Dabc][abcd] (see section 4.4 of the main paper).\footnote{These estimates are obtained with \textit{lem 1.0}. This does not provide an estimate of the Intercept, and *'s are produced when there is non-convergence.}

\begin{longtable}{l r r r r}
\hline
            & Estimate & Std. Error & z value & Pr \\
\hline
A          &   0.3479 &   0.8529 &    0.408  & 0.683 \\
B          &  -4.0938 &   0.7058 &   -5.801  & 0.000 \\
C          &   2.3301 &   0.8583 &    2.715  & 0.007 \\
D          &   0.2163 &   0.8624 &    0.251  & 0.802 \\
a          &  -5.6592 &   0.7150 &   -7.915  & 0.000 \\
b          & -10.2659 &   6.1800 &   -1.661  & 0.097 \\
c          &  -5.4417 &   0.0884 &  -61.578  & 0.000 \\
d          &  -3.7932 &   2.2503 &   -1.686  & 0.092 \\
AB         &   0.8199 &   0.6757 &    1.213  & 0.225 \\
AC         &   1.6865 &   0.8520 &    1.979  & 0.048 \\
BC         &   2.7625 &   0.7053 &    3.917  & 0.000 \\
AD         &  -0.5166 &   0.8584 &   -0.602  & 0.547 \\
BD         &  -1.5336 &   0.3480 &   -4.407  & 0.000 \\
CD         &  -0.0823 &   0.8629 &   -0.095  & 0.924 \\
Ba         &   2.0440 &   0.4367 &    4.681  & 0.000 \\
Ca         &  -0.7210 &   0.7350 &   -0.981  & 0.327 \\
Da         &  -4.9121 &   5.7005 &   -0.862  & 0.389 \\
Ab         &  -0.0194 &   0.0872 &   -0.222  & 0.824 \\
Cb         &   6.1502 &   6.1276 &    1.004  & 0.316 \\
Db         &  -0.2143 &   1.5289 &   -0.140  & 0.889 \\
ab         &  10.1675 &   5.4176 &    1.877  & 0.061 \\
Ac         &  -0.7919 &   0.0840 &   -9.422  & 0.000 \\
Bc         &   1.1438 &   0.0621 &   18.415  & 0.000 \\
Dc         &   0.0624 &   0.0802 &    0.778  & 0.436 \\
ac         &   3.9916 &   0.1232 &   32.397  & 0.000 \\
bc         &   3.7655 &   0.1740 &   21.637  & 0.000 \\
Ad         &  -7.6789 &   0.7041 &  -10.907  & 0.000 \\
Bd         &   1.4316 &   0.4690 &    3.052  & 0.002 \\
Cd         &  -0.1840 &   2.2659 &   -0.081  & 0.935 \\
ad         &   7.5486 &   5.0503 &    1.495  & 0.135 \\
bd         &   9.0030 &   2.3589 &    3.817  & 0.000 \\
cd         &   3.3894 &   0.1597 &   21.221  & 0.000 \\
ABC        &  -0.4778 &   0.6754 &   -0.707  & 0.479 \\
ABD        &   0.0706 &   0.0172 &    4.109  & 0.000 \\
ACD        &   0.5280 &   0.8589 &    0.615  & 0.539 \\
BCD        &  -0.8698 &   0.3421 &   -2.542  & 0.011 \\
BCa        &  -0.4423 &   0.4395 &   -1.006  & 0.314 \\
BDa        &   0.2749 &   0.4749 &    0.579  & 0.563 \\
CDa        &   5.8943 &   5.7036 &    1.033  & 0.301 \\
ACb        &  -0.3834 &   ****** &    *****  & 1.000 \\
ADb        &   0.5973 &   0.5116 &    1.168  & 0.243 \\
CDb        &   0.4713 &   1.5287 &    0.308  & 0.758 \\
Cab        &  -5.2224 &   5.3744 &   -0.972  & 0.331 \\
Dab        &   4.5710 &   4.7832 &    0.956  & 0.339 \\
ABc        &  -0.1932 &   0.0564 &   -3.427  & 0.001 \\
ADc        &   0.2368 &   0.0771 &    3.072  & 0.002 \\
BDc        &  -0.1576 &   0.0392 &   -4.020  & 0.000 \\
Bac        &  -0.2943 &   0.0552 &   -5.329  & 0.000 \\
Dac        &   0.0549 &   0.0593 &    0.927  & 0.354 \\
Abc        &   0.3765 &   0.1444 &    2.608  & 0.009 \\
Dbc        &   0.1462 &   0.1081 &    1.353  & 0.176 \\
abc        &  -1.9878 &   0.1860 &  -10.687  & 0.000 \\
ABd        &   0.0277 &   0.8817 &    0.031  & 0.975 \\
ACd        &   7.1994 &   0.6926 &   10.395  & 0.000 \\
BCd        &  -0.6969 &   0.4662 &   -1.495  & 0.135 \\
Bad        &  -1.9539 &   0.4836 &   -4.040  & 0.000 \\
Cad        &  -2.7067 &   5.0866 &   -0.532  & 0.595 \\
Abd        &   7.2026 &   0.1377 &   52.314  & 0.000 \\
Cbd        &  -6.2453 &   2.2712 &   -2.750  & 0.006 \\
abd        & -10.8600 &   0.1252 &  -86.768  & 0.000 \\
Acd        &   0.4996 &   0.1426 &    3.503  & 0.000 \\
Bcd        &  -0.3846 &   0.0910 &   -4.225  & 0.000 \\
acd        &  -2.0329 &   0.1822 &  -11.159  & 0.000 \\
bcd        &  -1.4389 &   0.2826 &   -5.091  & 0.000 \\
BCDa       &  -0.7668 &   0.4754 &   -1.613  & 0.107 \\
ACDb       &  -0.7367 &   0.5095 &   -1.446  & 0.148 \\
CDab       &  -5.2192 &   4.7859 &   -1.091  & 0.275 \\
ABDc       &  -0.0605 &   0.0258 &   -2.351  & 0.019 \\
BDac       &   0.0367 &   0.0391 &    0.938  & 0.348 \\
ADbc       &  -0.0837 &   0.0947 &   -0.884  & 0.377 \\
Dabc       &   0.0095 &   0.0871 &    0.109  & 0.913 \\
ABCd       &  -0.0569 &   0.8872 &   -0.064  & 0.949 \\
BCad       &   1.1125 &   0.4851 &    2.293  & 0.022 \\
ACbd       &  -6.7396 &   ****** &    *****  & 1.000 \\
Cabd       &   8.7740 &   ****** &    *****  & 1.000 \\
ABcd       &   0.0575 &   0.0738 &    0.779  & 0.436 \\
Bacd       &  -0.3435 &   0.0800 &   -4.291  & 0.000 \\
Abcd       &  -0.6356 &   0.2213 &   -2.871  & 0.004 \\
abcd       &   1.9426 &   0.2816 &    6.898  & 0.000 \\
  \hline
\end{longtable}

\newpage

\noindent Table S6: The parameter estimates from the four-register analysis using the restricted model [ABcd][AC][ADbc][BCad] [BDac][CDa][CDb][Abcd][Bacd][Dabc][abcd] (see section 4.4 of the main paper).

\begin{longtable}{l r r r r}
\hline
            & Estimate & Std. Error & z value & Pr \\
\hline
(Intercept) &  9.7463 &     NA       &  NA     & NA\\
A          &  0.2095 & 0.0232 &    9.0398 & 0.0000\\
B          & -3.4784 & 0.0294 & -118.1188 & 0.0000\\
c          & -5.4665 & 0.0897 &  -60.9648 & 0.0000\\
d          & -6.4289 & 0.1171 &  -54.8873 & 0.0000\\
C          &  2.1549 & 0.0239 &   90.0099 & 0.0000\\
D          & -0.5328 & 0.0311 &  -17.1521 & 0.0000\\
b          & -3.9471 & 0.1655 &  -23.8471 & 0.0000\\
a          & -5.9723 & 0.1765 &  -33.8387 & 0.0000\\
AB       &  0.3453 & 0.0065 &   53.1215 & 0.0000\\
Ac       & -0.7845 & 0.0874 &   -8.9779 & 0.0000\\
Bc       &  1.1475 & 0.0622 &   18.4443 & 0.0000\\
Ad       & -0.5200 & 0.0480 &  -10.8273 & 0.0000\\
Bd       &  2.8786 & 0.2282 &   12.6131 & 0.0000\\
cd       &  3.3103 & 0.1506 &   21.9822 & 0.0000\\
AC       &  1.8328 & 0.0229 &   80.1324 & 0.0000\\
AD       &  0.0155 & 0.0045 &    3.4757 & 0.0005\\
Ab       & -0.7716 & 0.0515 &  -14.9921 & 0.0000\\
Db       &  3.7140 & 0.1841 &   20.1739 & 0.0000\\
cD       &  0.1002 & 0.0788 &    1.2711 & 0.2037\\
cb       &  3.4215 & 0.1367 &   25.0334 & 0.0000\\
BC       &  2.2037 & 0.0289 &   76.1389 & 0.0000\\
Ba       &  1.4689 & 0.2587 &    5.6784 & 0.0000\\
Ca       & -0.4174 & 0.1734 &   -2.4067 & 0.0161\\
dC       &  2.5804 & 0.1019 &   25.3345 & 0.0000\\
da       &  7.0388 & 0.1579 &   44.5813 & 0.0000\\
BD       & -2.5067 & 0.0047 & -531.8307 & 0.0000\\
Da       & -2.4964 & 0.1614 &  -15.4655 & 0.0000\\
ca       &  3.9758 & 0.0862 &   46.1007 & 0.0000\\
CD       &  0.7595 & 0.0311 &   24.4071 & 0.0000\\
Cb       &  0.1679 & 0.1622 &    1.0351 & 0.3006\\
db       &  2.2739 & 0.1158 &   19.6379 & 0.0000\\
ba       &  4.8899 & 0.0581 &   84.1604 & 0.0000\\
ABc    & -0.1675 & 0.0551 &   -3.0381 & 0.0024\\
ABd    & -0.0119 & 0.0456 &   -0.2605 & 0.7945\\
Acd    &  0.4975 & 0.1448 &    3.4352 & 0.0006\\
Bcd    & -0.2914 & 0.0914 &   -3.1869 & 0.0014\\
ADb    & -0.2691 & 0.0445 &   -6.0485 & 0.0000\\
AcD    &  0.2463 & 0.0763 &    3.2274 & 0.0012\\
Acb    &  0.7104 & 0.1256 &    5.6543 & 0.0000\\
cDb    &  0.0298 & 0.1052 &    0.2832 & 0.7770\\
BCa    &  0.1615 & 0.2576 &    0.6270 & 0.5307\\
BdC    & -2.2512 & 0.2246 &  -10.0244 & 0.0000\\
Bda    & -2.9779 & 0.3733 &   -7.9782 & 0.0000\\
dCa    & -2.3192 & 0.1457 &  -15.9223 & 0.0000\\
BDa    & -0.5160 & 0.0188 &  -27.4183 & 0.0000\\
BcD    & -0.2554 & 0.0341 &   -7.4898 & 0.0000\\
Bca    & -0.2618 & 0.0565 &   -4.6333 & 0.0000\\
cDa    & -0.0568 & 0.0611 &   -0.9295 & 0.3526\\
CDa    &  3.4913 & 0.1591 &   21.9494 & 0.0000\\
CDb    & -3.3921 & 0.1840 &  -18.4329 & 0.0000\\
Adb    &  1.0287 & 0.1022 &   10.0607 & 0.0000\\
cdb    & -0.9482 & 0.2208 &   -4.2934 & 0.0000\\
cda    & -1.9031 & 0.1241 &  -15.3341 & 0.0000\\
Dba    & -0.5716 & 0.0498 &  -11.4843 & 0.0000\\
cba    & -1.9016 & 0.1255 &  -15.1524 & 0.0000\\
dba    & -2.0448 & 0.1049 &  -19.4944 & 0.0000\\
ABcd &  0.0149 & 0.0750 &    0.1992 & 0.8421\\
AcDb &  0.0337 & 0.0915 &    0.3679 & 0.7130\\
BdCa &  2.2211 & 0.3715 &    5.9793 & 0.0000\\
BcDa &  0.0940 & 0.0400 &    2.3523 & 0.0187\\
Acdb & -1.1363 & 0.1968 &   -5.7739 & 0.0000\\
Bcda & -0.4754 & 0.0825 &   -5.7636 & 0.0000\\
cDba &  0.0517 & 0.0901 &    0.5743 & 0.5658\\
cdba &  1.8568 & 0.1911 &    9.7184 & 0.0000\\            
\hline
\end{longtable}

\newpage

\noindent\textbf{S3: Code}
\vspace{5mm}

\noindent All code is openly available at \url{https://github.com/MaartenCruyff/cenzus}. Here the data are available as well.